\documentclass{elsarticle}

\usepackage[reviewcopy]{adndt}
\usepackage{longtable}

\usepackage{amsmath}

\usepackage{amssymb}

\biboptions{square,sort&compress}
\bibpunct[]{[}{]}{,}{n}{}{;}
\begin{document}

\begin{frontmatter}

\journal{Atomic Data and Nuclear Data Tables}


\title{Discovery of dysprosium, holmium, erbium, thulium and ytterbium isotopes}

\author{C. Fry}
\author{M. Thoennessen\corref{cor1}}\ead{thoennessen@nscl.msu.edu}

 \cortext[cor1]{Corresponding author.}

 \address{National Superconducting Cyclotron Laboratory and \\ Department of Physics and Astronomy, Michigan State University, \\ East Lansing, MI 48824, USA}

\begin{abstract}
Currently, 31 dysprosium, 32 holmium, 32 erbium, 33 thulium and 31 ytterbium isotopes have been observed and the discovery of these isotopes is discussed here. For each isotope a brief synopsis of the first refereed publication, including the production and identification method, is presented.
\end{abstract}

\end{frontmatter}





\newpage
\tableofcontents
\listofDtables

\vskip5pc

\section{Introduction}\label{s:intro}

The discovery of dysprosium, holmium, erbium, thulium and ytterbium isotopes is discussed as part of the series summarizing the discovery of isotopes, beginning with the cerium isotopes in 2009 \cite{2009Gin01}. Guidelines for assigning credit for discovery are (1) clear identification, either through decay-curves and relationships to other known isotopes, particle or $\gamma$-ray spectra, or unique mass and Z-identification, and (2) publication of the discovery in a refereed journal. The authors and year of the first publication, the laboratory where the isotopes were produced as well as the production and identification methods are discussed. When appropriate, references to conference proceedings, internal reports, and theses are included. When a discovery includes a half-life measurement the measured value is compared to the currently adopted value taken from the NUBASE evaluation \cite{2003Aud01} which is based on the ENSDF database \cite{2008ENS01}. In cases where the reported half-life differed significantly from the adopted half-life (up to approximately a factor of two), we searched the subsequent literature for indications that the measurement was erroneous. If that was not the case we credited the authors with the discovery in spite of the inaccurate half-life. All reported half-lives inconsistent with the presently adopted half-life for the ground state were compared to isomers half-lives and accepted as discoveries if appropriate following the criterium described above.

The first criterium excludes measurements of half-lives of a given element without mass identification. This affects mostly isotopes first observed in fission where decay curves of chemically separated elements were measured without the capability to determine their mass. Also the four-parameter measurements (see, for example, Ref. \cite{1970Joh01}) were, in general, not considered because the mass identification was only $\pm$1 mass unit.

The second criterium affects especially the isotopes studied within the Manhattan Project. Although an overview of the results was published in 1946 \cite{1946TPP01}, most of the papers were only published in the Plutonium Project Records of the Manhattan Project Technical Series, Vol. 9A, ``Radiochemistry and the Fission Products,'' in three books by Wiley in 1951 \cite{1951Cor01}. We considered this first unclassified publication to be equivalent to a refereed paper.

The initial literature search was performed using the databases ENSDF \cite{2008ENS01} and NSR \cite{2008NSR01} of the National Nuclear Data Center at Brookhaven National Laboratory. These databases are complete and reliable back to the early 1960's. For earlier references, several editions of the Table of Isotopes were used \cite{1940Liv01,1944Sea01,1948Sea01,1953Hol02,1958Str01,1967Led01}. A good reference for the discovery of the stable isotopes was the second edition of Aston's book ``Mass Spectra and Isotopes'' \cite{1942Ast01}.

\section{Discovery of $^{139-169}$Dy}
Thirty-one dysprosium isotopes from A = 139$-$169 have been discovered so far; these include 7 stable, 19 neutron-deficient and 5 neutron-rich isotopes. According to the HFB-14 model \cite{2007Gor01}, on the neutron-rich side the bound isotopes should reach at least up to $^{217}$Dy while on the neutron deficient side five more isotopes should be particle stable ($^{134-138}$Dy). Five additional isotopes ($^{129-133}$Dy) could still have half-lives longer than 10$^{-9}$~s \cite{2004Tho01}. Thus, about 58 isotopes have yet to be discovered corresponding to 68\% of all possible dysprosium isotopes.

\begin{figure}
	\centering
	\includegraphics[scale=.7]{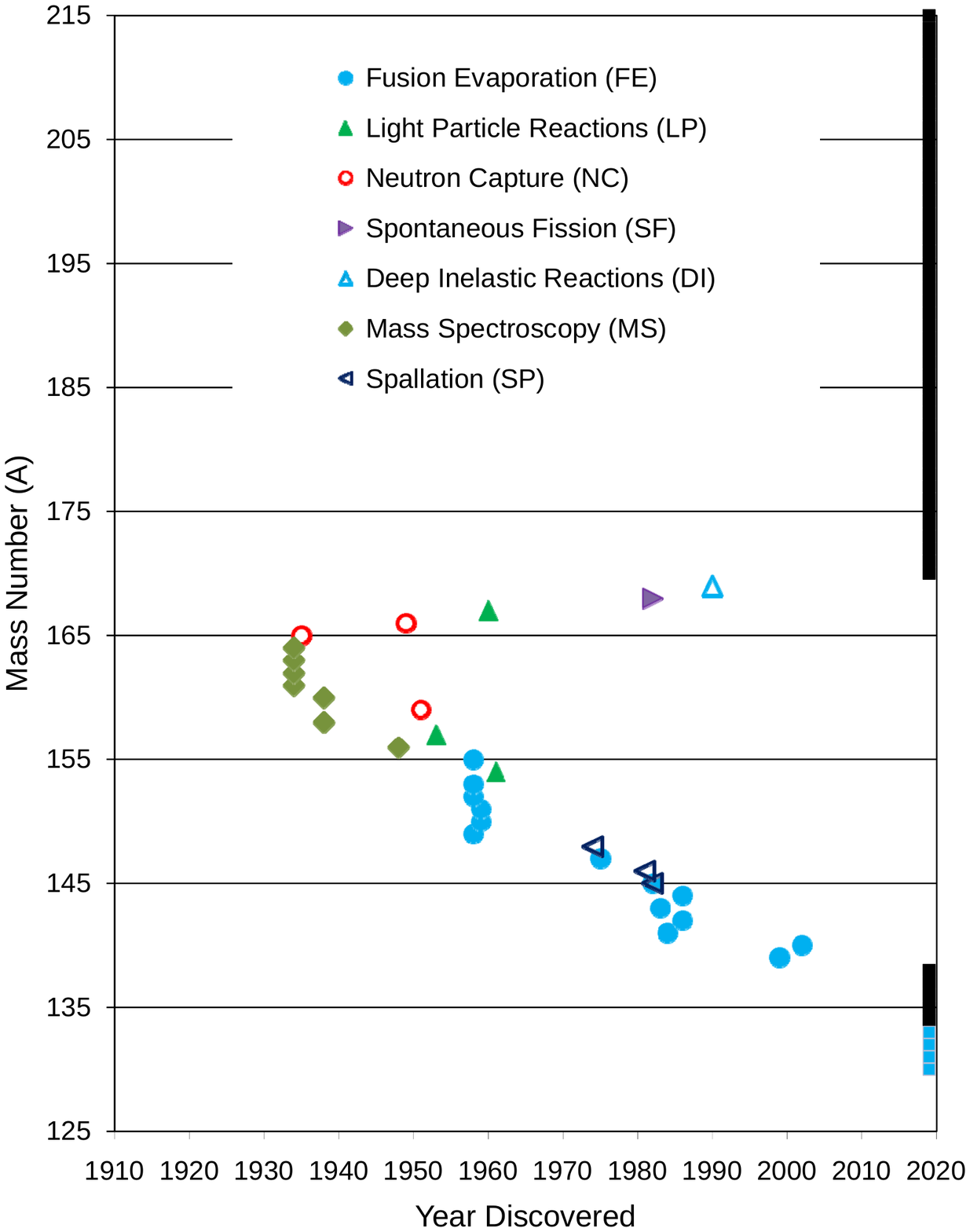}
	\caption{Dysprosium isotopes as a function of time when they were discovered. The different production methods are indicated. The solid black squares on the right hand side of the plot are isotopes predicted to be bound by the HFB-14 model. On the proton-rich side the light blue squares correspond to unbound isotopes predicted to have half-lives larger than $\sim 10^{-9}$~s.}
\label{f:year-dy}
\end{figure}

Figure \ref{f:year-dy} summarizes the year of discovery for all dysprosium isotopes identified by the method of discovery. The range of isotopes predicted to exist is indicated on the right side of the figure. The radioactive dysprosium isotopes were produced using fusion evaporation reactions (FE), light-particle reactions (LP), deep-inelastic reactions (DI), spontaneous fission (SF), neutron capture (NC), and spallation (SP). The stable isotopes were identified using mass spectroscopy (MS). The discovery of each dysprosium isotope is discussed in detail and a summary is presented in Table 1.

\subsection*{$^{139}$Dy}\vspace{0.0cm}
Xu et al. first identified $^{139}$Dy in 1999 and reported the results in ``New $\beta$-delayed proton precursors in the rare-earth region near the proton drip line'' \cite{1999Xu01}. A 176 MeV $^{36}$Ar beam was accelerated with the Lanzhou sector-focused cyclotron and bombarded an enriched $^{106}$Cd target. Proton-$\gamma$ coincidences were measured in combination with a He-jet type transport system. ``A clear 221-keV $\gamma$ peak and a tiny 384-keV $\gamma$ peak in the proton-coincident $\gamma$(x)-ray spectrum in the $^{36}$Ar+$^{106}$Cd reaction were assigned to the 2$^+\rightarrow$0$^+$ and 4$^+\rightarrow$2$^+$ $\gamma$ transitions in the `daughter' nucleus $^{138}$Gd of the $\beta p$ precursor $^{139}$Dy.'' The observed half-life of 0.6(2)~s corresponds to the currently accepted value.

\subsection*{$^{140}$Dy}\vspace{0.0cm}
Krolas et al. observed $^{140}$Dy as reported in the 2002 article, ``First observation of the drip line nucleus $^{140}$Dy: Identification of a 7 $\mu$s K isomer populating the ground state band'' \cite{2002Kro01}. A 315 MeV $^{54}$Fe beam, accelerated by the Oak Ridge tandem accelerator, bombarded an enriched $^{92}$Mo target. Fusion-evaporation products were separated with the RMS (Recoil Mass Spectrometer) and detected in a microchannel plate detector in coincidence with $\gamma$-rays in the Clover Germanium Detector Array for Recoil Decay Spectroscopy CARDS. ``A new 7 $\mu$s isomer in the drip line nucleus $^{140}$Dy was selected from the products of the $^{54}$Fe (315 MeV)+$^{192}$Mo reaction by a recoil mass spectrometer and studied with recoil-delayed $\gamma-\gamma$ coincidences.'' Less than a month later Cullen et al.\ independently reported excited states in $^{140}$Dy \cite{2002Cul01}.

\subsection*{$^{141}$Dy}\vspace{0.0cm}
In the 1984 article ``Beta-delayed proton emission observed in new lanthanide isotopes'' Nitschke et al. reported the first observation of $^{141}$Dy \cite{1984Nit01}. A 274 MeV $^{54}$Fe beam from the Berkeley SuperHILAC was used to form $^{141}$Dy in the fusion-evaporation reaction $^{92}$Mo($^{54}$Fe,$\alpha$n). Beta-delayed protons and characteristic X-rays were measured in coincidence at the on-line isotope separator OASIS. ``The ensemble of these observations lead us to the conclusion that two new beta delayed proton emitters $^{141}$Dy and $^{141}$Gd were being observed''. The measured half-life of 1.0(2)~s is in agreement with the currently accepted value of 0.9(2)~s.

\subsection*{$^{142}$Dy}\vspace{0.0cm}
The first observation of $^{142}$Dy was reported by Wilmarth et al. in their 1986 paper entitled ``Beta-delayed proton emission in the lanthanide region'' \cite{1986Wil01}. A 247 MeV $^{54}$Fe beam from the Berkeley Super HILAC bombarded a $^{92}$Mo target and $^{142}$Dy was produced in the fusion-evaporation reaction $^{92}$Mo($^{54}$Fe,2p2n). Beta-delayed particles, X-rays and $\gamma$-rays were measured following mass separation with the on-line isotope separator OASIS. ``A $\beta$-delayed proton activity with a single component half-life of 1.8$\pm$0.6~s is assigned to the new isotope $^{142}$Dy, on the basis of Tb K x-rays observed in coincidence with protons. Independent analysis of the decay of the Tb x-rays and associated $\gamma$-rays results in a $^{142}$Dy half-life of 2.3$\pm$0.8~s.'' This half-life agrees with the accepted value 2.3(3)~s.

\subsection*{$^{143}$Dy}\vspace{0.0cm}
In the 1983 article ``Beta-delayed proton emission observed in new lanthanide isotopes'' Nitschke et al. reported the first observation of $^{143}$Dy \cite{1983Nit01}. A $^{92}$Mo target was bombarded with 275 MeV $^{56}$Fe and 292 MeV $^{58}$Ni from the Berkeley SuperHILAC and $^{143}$Dy was formed in the fusion-evaporation reactions $^{92}$Mo($^{56}$Fe,$\alpha$n) and $^{92}$Mo($^{58}$Ni,$\alpha$2pn), respectively. Beta-delayed protons and characteristic X-rays were measured in coincidence at the on-line isotope separator OASIS. ``In two experiments with $^{58}$Ni and $^{56}$Fe beams on $^{92}$Mo targets, beta-delayed protons with similar half-lives and similar energy distributions were observed. The weighted average half-life of 4.1$\pm$0.3~s is in agreement with the calculated value of 3.2~s for $^{143}$Dy.''. This half-life corresponds to the currently accepted value.

\subsection*{$^{144}$Dy}\vspace{0.0cm}
Redon et al. described the first observation of $^{144}$Dy in the 1986 paper ``Exotic neutron-deficient nuclei near N=82'' \cite{1986Red01}. Enriched $^{112}$Sn targets were bombarded with a 191 MeV $^{35}$Cl beam from the Grenoble SARA accelerator and $^{144}$Dy was formed in the fusion evaporation residue reaction $^{112}$Sn($^{35}$Cl,p2n). The residues were separated with an on-line mass separator and a He-jet system. X-ray and $\gamma$-ray spectra were measured. ``The 196 and 298 keV $\gamma$-rays decay with T$_{1/2}$ = 9.0$\pm$0.7~s (Takahashi prediction = 7~s) and therefore belong to the $^{144}$Dy $\beta$-decay''. This half-life is used in the calculation of the currently accepted value. Three months later a 7(3)~s half-life was reported for $^{144}$Dy independently by Wilmarth et al.\ \cite{1986Wil01}.

\subsection*{$^{145}$Dy}\vspace{0.0cm}
In 1982, $^{145}$Dy was simultaneously discovered by Nolte et al. in ``Very proton rich nuclei with N$\approx$82'' \cite{1982Nol02} and by Alkhazov et al. in ``New neutron deficient isotopes with mass numbers A=136 and 145'' \cite{1982Alk01}. Nolte et al. used the Munich tandem and heavy-ion linear rf post accelerator to bombard $^{90}$Zr with 233$-$250~MeV $^{58}$Ni and $^{145}$Dy was populated in the fusion-evaporation reaction $^{90}$Zr($^{58}$Ni,n2p). Gamma-ray singles and coincidence spectra were measured with Ge(Li) detectors. ``From this fit, the half-life of $^{145}$Dy was determined to be 13.6$\pm$1~s.'' Alkhazov et al. populated $^{145}$Dy by spallation of 1 GeV protons on tungsten and tantalum targets. X-rays, $\gamma$-rays and conversion electrons were measured following mass separation with the IRIS on-line mass separator facility. ``The new gamma lines with the following energies and intensities: 39.7 /?/, 578.2 /100/, 639.6 /93/ and 804.3 /77/ and with the accepted half-life T$_{l/2}$=18$\pm$3~s belong to the decay of $^{145}$Dy.'' These half-lives are consistent with the presently adopted value of 14.1(7)~s for an isomeric state. A month earlier Gui et al. had reported the observation of $\gamma$-rays in $^{145}$Dy without giving any details \cite{1982Gui01}.

\subsection*{$^{146}$Dy}\vspace{0.0cm}
Alkhazov et al. identified $^{146}$Dy in 1981 in ``New isotope $^{146}$Dy'' \cite{1981Alk03}. A tungsten target was bombarded with 1 GeV protons from the Leningrad synchrocyclotron. X-rays and $\gamma$-ray spectra were measured with Ge(Li) detectors following mass separation with the IRIS on-line mass separator facility. ``The second set we ascribe to the decay $^{146}$Dy$\rightarrow ^{146}$Tb. The analysis of the decay data for K$_{\alpha_1}$ Tb line, gives T$_{1/2}$ = 31$\pm$5~s, which is in good agreement with the value predicted by Tokahashi et al. for the half-life of $^{146}$Dy.'' This value agrees with the currently accepted value of 29(3)~s.

\subsection*{$^{147}$Dy}\vspace{0.0cm}
In 1975, the discovery of $^{147}$Dy was announced in the paper ``Excitation energies of the h$_{11/2}$ and d$_{3/2}$ neutron states in $^{145}$Gd and $^{147}$Dy'' by Toth et al. \cite{1975Tot02}. A $^{141}$Pr target was bombarded with 124$-$157~MeV $^{14}$N beams from the Oak Ridge isochronous cyclotron. A capillary transport system extracted the product nuclei to a shielded area where singles and coincidence $\gamma$-ray measurements were taken with Ge(Li) detectors. ``A systematic shift in the X-ray energies can be noted by comparing the three sets of spectra. Based on this and other evidence the 72 and 679 keV $\gamma$-rays are assigned to $^{147}$Dy$^{m}$''. Three weeks earlier Firestone et al.\ had mentioned a 72~keV $\gamma$-ray in $^{147}$Dy without a reference \cite{1975Fir01}.

\subsection*{$^{148}$Dy}\vspace{0.0cm}
``Method for obtaining separated short-lived isotopes of rare earth elements'' was published in 1974 by Latuszynski et al.\ documenting their observation of $^{148}$Dy \cite{1974Lat02}. A tantalum target was bombarded with 660~MeV protons from the Dubna synchrocyclotron. Gamma-ray spectra and decay curves were measured at the end of an electromagnetic separator. ``Using the method proposed for investigations in the field of nuclear spectroscopy the gamma-spectra of short-living isotopes with T$_{1/2} \le$ 1 minute have been measured. The new isotopes $^{161}$Yb (4.2~min), $^{148}$Dy (3.5~min) $^{132}$Pr (1.6~min) have been identified.'' The observed half-life is used in the calculation of the currently accepted value of 3.3(2)~min.

\subsection*{$^{149}$Dy}\vspace{0.0cm}
In 1958, Toth and Rassmussen observed $^{149}$Dy as described in ``Studies of rare earth alpha emitters'' \cite{1958Tot02}. Praseodymium was bombarded with a beam of $^{14}$N from the Berkeley 60-in.\ cyclotron. Decay curves of the subsequent activities were measured. ``The Tb$^{149}$ growth curve indicates the presence of Dy$^{149}$, which decays to Tb$^{149}$. The growing-in period and the 4-hr tail are shown in [the figure], and from this curve we determine a half-life of 8$\pm$2 min for Dy$^{149}$.'' This half-life is almost a factor of two larger than the currently accepted value of 4.20(14)~min. This discrepancy was later explained by Bingham et al.\ \cite{1973Bin01}.

\subsection*{$^{150,151}$Dy}\vspace{0.0cm}
``$\gamma$-rays following the $\beta$-decay of rare-earth $\alpha$-emitters'' was published by Toth and Rasmussen in 1959, describing the discovery of $^{150}$Dy and $^{151}$Dy \cite{1959Tot03}. $^{14}$N was accelerated to 140~MeV by the Berkeley heavy ion accelerator to bombard a praseodymium target forming $^{150}$Dy and $^{151}$Dy in (5n) and (4n) fusion-evaporation reactions, respectively. Alpha-particle decay curves and $\gamma$-ray spectra were measured. ``In addition, the mass assignments of $^{150}$Dy and $^{151}$Dy, the 8 min and 19 min dysprosium $\alpha$-emitters respectively, have been determined with certainty.''  These half-lives agree with the presently adopted values of 7.17(5)~min and 17.9(3)~min, for $^{150}$Dy and $^{151}$Dy, respectively. Previously, Rasmussen et al. had assigned half-lives of 7(2)~min and 19(4)~min to dysprosium isotopes with 149$\le$A$\le$153 \cite{1953Ras01}.

\subsection*{$^{152,153}$Dy}\vspace{0.0cm}
Toth and Rasmussen reported the discoveries of $^{152}$Dy and $^{153}$Dy in the 1958 paper ``Studies of rare earth alpha emitters'' \cite{1958Tot02}. $^{152}$Gd was bombarded with 48~MeV alpha particles accelerated by the Berkeley 60-in.\ cyclotron. Subsequent $\alpha$ emission was measured following chemical separation. The mass assignment was achieved with the stacked foil technique. ``The 2.5-hr isotope was seen in the first target foil at 48 Mev but was absent at 41.5 Mev. At least, it was not present in a sufficient amount to be noticed. The ($\alpha$,4n) reactions, in this region, have thresholds at about 39 Mev. It seems quite reasonable to assume then that the activity was produced by an ($\alpha$,4n) reaction on Gd$^{162}$ and must be Dy$^{162}$. The 5-hr isotope was present only in a small amount at 33.5 Mev and was absent at 23.3 Mev. Since ($\alpha$,3n) thresholds are approximately at 28 Mev, one is forced to the conclusion that the 5-hr alpha emitter must have been made by an ($\alpha$,3n) reaction on Gd$^{162}$. It must be Dy$^{163}$.'' The quoted half-lives of 2.3(2)~h ($^{152}$Dy) and 5.0(5)~h ($^{153}$Dy) are consistent with the currently accepted values of 2.38(2)~h and 6.4(1)~h, respectively. Previously, Rasmussen et al. had assigned the half-life of 2.3(2)~h to a dysprosium isotope with 149$\le$A$\le$153 \cite{1953Ras01}.

\subsection*{$^{154}$Dy}\vspace{0.0cm}
The discovery of $^{154}$Dy was reported in the 1961 paper ``Dysprosium-154, a long-lived $\alpha$-emitter'' by Macfarlane \cite{1961Mac01}. Gadolinium oxide samples enriched in $^{154}$Gd were irradiated with 48~MeV and 37~MeV $\alpha$ particles from the Berkeley 60~in cyclotron. Alpha spectra were obtained using a Frisch-grid argon-methane flow-type ion chamber following chemical separation. ``From these data, the alpha half-life of $^{154}$Dy was determined to be 1$\times$10$^{6}$~years, uncertain by a factor of three.'' This value is in agreement with the currently accepted half-life 3.0$\times$10$^{6}$~y. A previously reported half-life of 13(2)~h \cite{1958Tot02} was evidently incorrect.

\subsection*{$^{155}$Dy}\vspace{0.0cm}
Toth and Rasmussen reported the discovery of $^{155}$Dy in the 1958 paper ``Studies of rare earth alpha emitters'' \cite{1958Tot02}. Natural gadolinium and enriched $^{154}$Gd were bombarded with 48~MeV alpha particles from the Berkeley 60-in. cyclotron. $^{155}$Dy was produced by ($\alpha$,4n) and ($\alpha$,3n) reactions on $^{155}$Gd and $^{154}$Gd, respectively. ``Our mass assignment of Dy$^{155}$ was accomplished in the following manner: This new isotope has a prominent gamma transition of 225 kev which was found to decay with a 10-hr half-life.''  This half-life agrees with the currently accepted value of 9.9(2)~h. A previously reported half-life of 20~h \cite{1957Mih01} was evidently incorrect \cite{1958Dob01}.

\subsection*{$^{156}$Dy}\vspace{0.0cm}
``A new naturally occurring isotope of dysprosium'' was published in 1948 by Hess and Inghram describing the observation of $^{156}$Dy \cite{1948Hes01}. Dysprosium oxide samples were placed on a surface ionization type filament of a mass spectrometer. Three separate samples were investigated using optical spectrographic analysis at Argonne. ``However, in spite of the known presence of all these impurities and consideration of other possible impurities, the peak observed at mass 156 in the metallic ion group could not be explained... We therefore conclude that dysprosium has a previously unknown isotope of mass 156 which is present to about 0.05 percent of the total.''

\subsection*{$^{157}$Dy}\vspace{0.0cm}
Handley and Olson discovered $^{157}$Dy as reported in the 1953 paper ``Dysprosium 157'' \cite{1953Han02}. Terbium oxide was bombarded with 24 Mev protons from the Oak Ridge 86-in.\ cyclotron. Decay curves and $\gamma$-spectra were measured following chemical separation. An excitation energy function was measured with the stacked foil technique. ``From the [the figure] the high threshold of 19$\pm$1 Mev indicates a (p,3n) reaction, thus assigning the 8.2-hour activity to Dy$^{157}$.''  This half-life is included in the calculation of the current value of 8.14(4)~h.

\subsection*{$^{158}$Dy}\vspace{0.0cm}
$^{158}$Dy was discovered by Dempster as described in the 1938 paper ``The isotopic constitution of gadolinium, dysprosium, erbium and ytterbium'' \cite{1938Dem01}. Two samples of dysprosium oxide reduced with lanthanum and another sample reduced in addition with calcium were used for the analysis in the Chicago mass spectrograph. ``Both samples showed two new isotopes at masses 160 and 158... The isotope at 160 was found on 12 photographs with exposures from one-half to forty minutes, the one at 158 on 4 photographs''.

\subsection*{$^{159}$Dy}\vspace{0.0cm}
Butement described the observation of $^{159}$Dy  in ``Radioactive $^{159}$Dy'' in 1951 \cite{1951But02}.  A sample of dysprosium oxide was irradiated with neutrons from the Harwell pile. Decay curves and X-rays were measured following chemical separation. $^{159}$Dy was also produced in the reaction $^{159}$Tb(d,2n). ``After waiting 60 days for complete decay of the $^{166}$Dy, the decay of the long-lived activity was followed for 400 days, and showed a half-life of 132 days... It is concluded that the only long-lived product of neutron capture by dysprosium is $^{159}$Dy.'' This half-life agrees with the currently accepted value of 144.4(2)~d. Butement had mentioned the observation of a half-life $>$50~days of $^{159}$Dy in an earlier paper \cite{1950But04}. Previously, a 140(10)~d half-life was assigned to either $^{157}$Dy or $^{159}$Dy \cite{1949Ket01}.

\subsection*{$^{160}$Dy}\vspace{0.0cm}
$^{160}$Dy was discovered by Dempster as described in the 1938 paper ``The isotopic constitution of gadolinium, dysprosium, erbium and ytterbium'' \cite{1938Dem01}. Two samples of dysprosium oxide reduced with lanthanum and another sample reduced in addition with calcium were used for the analysis in the Chicago mass spectrograph. ``Both samples showed two new isotopes at masses 160 and 158... The isotope at 160 was found on 12 photographs with exposures from one-half to forty minutes, the one at 158 on 4 photographs''.

\subsection*{$^{161-164}$Dy}\vspace{0.0cm}
In 1934, Aston reported the first observation of stable $^{161}$Dy, $^{162}$Dy, $^{163}$Dy, and $^{164}$Dy in ``The isotopic constitution and atomic weights of the rare earth elements'' \cite{1934Ast04}. Rare earth elements were analyzed with the Cavendish mass spectrograph. ``Dysprosium (66) gave poor spectra but sufficient to indicate that it consists of mass numbers 161, 162, 163, 164 not differing much in relative abundance.''

\subsection*{$^{165}$Dy}\vspace{0.0cm}
In 1935, $^{165}$Dy was discovered simultaneously by Marsh and Sugden in ``Artificial radioactivity of the rare earth elements'' \cite{1935Mar01} and Hevesy and Levi in ``Artificial radioactivity of dysprosium and other rare earth elements'' \cite{1935Hev01} published back-to-back in the same issue of Nature. Marsh and Sugden irradiated dysprosium oxide with neutrons from a 400 mCi radon source in contact with powdered beryllium. In the table a half-life of 2.5(1)~h is listed for $^{165}$Dy. Hevesy and Levi bombarded dysprosium oxide with neutrons from a few hundred mCi radium emanation source. ``We find that dysprosium shows an unusually strong activity due to $_{66}$Dy$^{165}$ under the action of slow neutrons; so far as we can ascertain, it is the strongest activity found hitherto.'' In a table a half-life of 2.5~h is listed. These half-lives agree with the presently adopted value of 2.334(1)~h.

\subsection*{$^{166}$Dy}\vspace{0.0cm}
In the 1949 paper ``New radioactive isotopes of dysprosium'' Ketelle reported the observation of $^{166}$Dy \cite{1949Ket01}. Slow neutrons from the Oak Ridge reactor were used to irradiate dysprosium oxide. Decay curves and absorption spectra were recorded following chemical separation. ``Since both the half-life and the energy of the daughter activity agree with those of Ho$^{166}$, we conclude that the 80-hr.\ parent is Dy$^{166}$.'' This half-life agrees with the currently accepted value of 81.6(1)~h. Just over a month later Butement independently reported a half-life of 82~h \cite{1950But04}.

\subsection*{$^{167}$Dy}\vspace{0.0cm}
$^{167}$Dy was first observed by Wille and Fink as reported in the 1960 paper ``Activation cross sections for 14.8-Mev neutrons and some new radioactive nuclides in the rare earth region'' \cite{1960Wil02}. An enriched $^{170}$Er sample was irradiated with neutrons produced from the $^3$H(d,n)$^4$He reaction from the Arkansas Cockcroft-Walton accelerator. Decay curves were measured with an aluminum-walled, methane-flow beta-proportional counter. ``Therefore, it seems likely that the 40-sec activity is Ho$^{170}$ from the Er$^{170}$(n,p) reaction and the 4.4-min period is Dy$^{167}$ from the Er$^{170}$(n,$\alpha$) reaction.''  This half-life is within a factor of two of the currently accepted value of 6.20(8)~min.

\subsection*{$^{168}$Dy}\vspace{0.0cm}
In the 1982 paper ``Identification of a new isotope, $^{168}$Dy'' Gehrke et al. described the discovery of $^{168}$Dy \cite{1982Geh01}. Fission products from spontaneous fission of $^{252}$Cf were chemically separated and multiscaled $\gamma$-ray spectra were measured with a Ge(Li) detector. ``A previously unreported nuclide, $^{168}$Dy, has been identified and found to have a halflife of 8.5$\pm$0.5 min.'' This half-life is included in the calculation of the currently accepted value of 8.7(3)~min.

\subsection*{$^{169}$Dy}\vspace{0.0cm}
Chasteler et al.\ observed $^{169}$Dy and published the results in their 1990 paper ``Decay of the neutron-rich isotope $^{171}$Ho and the identification of $^{169}$Dy'' \cite{1990Cha01}). The Berkeley SuperHILAC was used to bombard a natural tungsten target with a 8.5~MeV/u $^{170}$Er beam and $^{169}$Dy was produced in multinucleon transfer reactions. Beta- and gamma-ray spectra were measured following on-line mass separation with the OASIS facility. ``Based on the agreement between both the experimental decay energy and half-life with the predictions, we assign the observed activity to the new isotope $^{169}$Dy with a half-life of 39(8)~s as determined from the weighted average of both the $\gamma$ and $\beta$ half-lives.'' This half-life is the currently accepted value.


\section{Discovery of $^{140-172}$Ho}
Thirty-two holmium isotopes from A = 140$-$172 have been discovered so far; these include 1 stable, 24 neutron-deficient and 7 neutron-rich isotopes. According to the HFB-14 model \cite{2007Gor01}, on the neutron-rich side the bound isotopes should reach at least up to $^{223}$Ho while on the neutron deficient side the dripline has been crossed with the observation of the proton emitters $^{140}$Ho and $^{141}$Ho. Five additional isotopes ($^{135-139}$Ho) could still have half-lives longer than 10$^{-9}$~s \cite{2004Tho01}. Thus, about 57 isotopes have yet to be discovered corresponding to 63\% of all possible holmium isotopes.

\begin{figure}
	\centering
	\includegraphics[scale=.7]{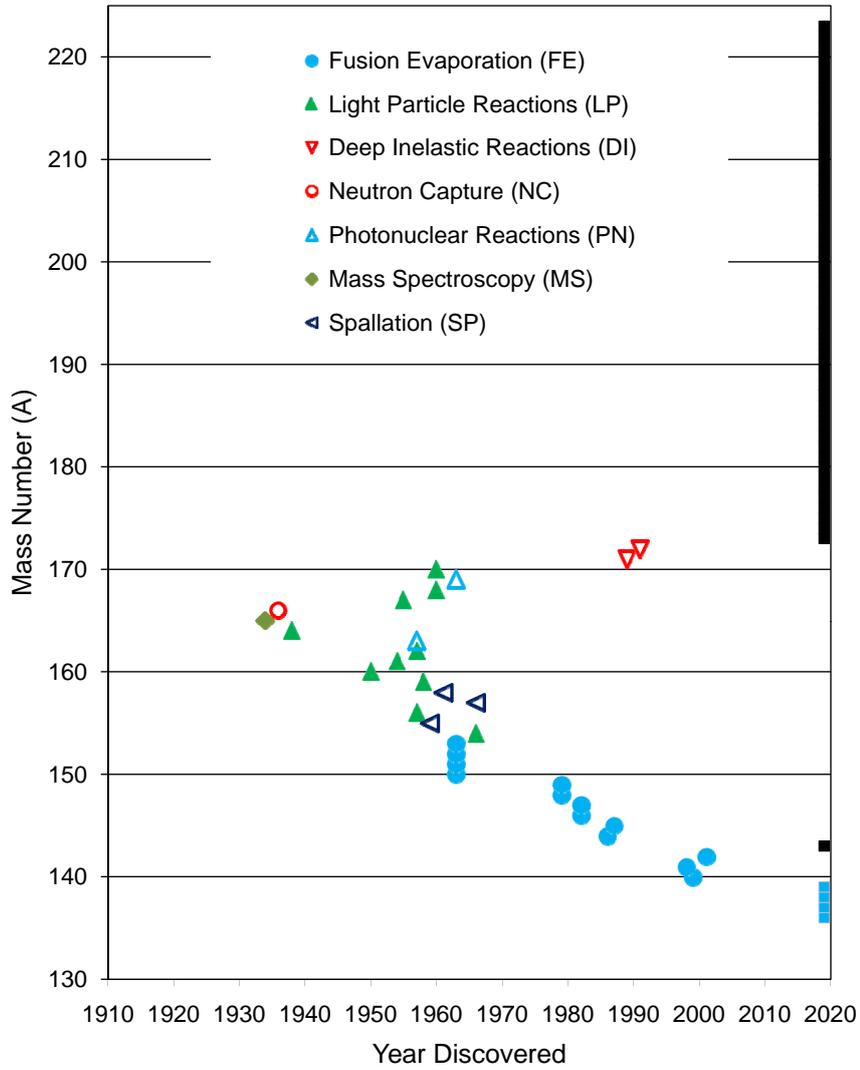}
	\caption{Holmium isotopes as a function of time when they were discovered. The different production methods are indicated. The solid black squares on the right hand side of the plot are isotopes predicted to be bound by the HFB-14 model. On the proton-rich side the light blue squares correspond to unbound isotopes predicted to have half-lives larger than $\sim 10^{-9}$~s.}
\label{f:year-ho}
\end{figure}

Figure \ref{f:year-ho} summarizes the year of discovery for all holmium isotopes identified by the method of discovery. The range of isotopes predicted to exist is indicated on the right side of the figure. The radioactive holmium isotopes were produced using fusion evaporation reactions (FE), light-particle reactions (LP), deep-inelastic reactions (DI), photo-nuclear reactions (PN), neutron capture (NC), and spallation (SP). The stable isotope was identified using mass spectroscopy (MS). The discovery of each holmium isotope is discussed in detail and a summary is presented in Table 1.

\subsection*{$^{140}$Ho}\vspace{0.0cm}
The discovery of $^{140}$Ho was reported in ``Proton emitters $^{140}$Ho and $^{141}$Ho: Probing the structure of unbound Nilsson orbitals'' by Rykaczewski et al.\ in 1999 \cite{1999Ryk01}. An enriched $^{92}$Mo target was bombarded with 315-MeV $^{54}$Fe at the Holifield Radioactive Ion Beam Facility at Oak Ridge. Reaction products were separated with the RMS recoil mass separator, identified with a position-sensitive avalanche counter and implanted in double-sided silicon strip detector DSSD. The DSSD also measured time-correlated decay events. ``Two new proton emitting states in the deformed nuclei $^{140}$Ho (E$_{p}$=1086$\pm$10~keV, T$_{1/2}$ = 6$\pm$3~ms) and $^{141m}$Ho (E$_{p}$=1230$\pm$20~keV, T$_{1/2}$=8$\pm$3~$\mu$s) have been identified.''  The quoted half-life is the currently accepted value.

\subsection*{$^{141}$Ho}\vspace{0.0cm}
Davids et al.\ observed $^{141}$Ho in 1998 and published their results in ``Proton radioactivity from highly deformed nuclei'' \cite{1998Dav01}. $^{54}$Fe beams accelerated to 285 and 305 MeV by the Argonne ATLAS accelerator bombarded a $^{92}$Mo target and $^{141}$Ho was formed in the fusion evaporation reaction $^{92}$Mo($^{54}$Fe,p4n). Reaction products were separated with the Fragment Mass Analyzer and implanted in a double-sided silicon strip detector where subsequent protons were recorded.  ``The low decay energy rules out $\alpha$ radioactivity, and we assign this peak to proton radioactivity from $^{141}$Ho, produced with a cross section $\sigma \sim$ 250 nb (at both beam energies).'' The observed half-life of 4.2(4)~ms is used in the calculation of the currently accepted value of 4.1(3)~ms.

\subsection*{$^{142}$Ho}\vspace{0.0cm}
In 2001 Xu et al.\ described the observation of $^{142}$Ho in ``$\beta$-delayed proton decay of the proton drip-line nucleus $^{142}$Ho'' \cite{2001Xu02}. The Lanzhou sector-focusing cyclotron was used to bombard enriched $^{106}$Cd targets with a 232-MeV $^{40}$Ca beam and $^{142}$Ho was formed in the (p3n) fusion-evaporation reaction. A combination of a He jet and tape system transported the reaction products to a counting station. Protons were measured with silicon surface barrier detectors and $\gamma$- and X-rays were detected with a coaxial HpGe(GMX) detector. ``The decay curve of the 307-keV $\gamma$ line coincident with 2.5-6.5 MeV protons, from which the half-life of the new nuclide $^{142}$Ho was extracted to be 0.4$\pm$0.1~s, is shown in the inset of [the figure].'' The quoted half-life corresponds to the currently accepted value.

\subsection*{$^{144}$Ho}\vspace{0.0cm}
The first observation of $^{144}$Ho was reported by Wilmarth et al.\ in their 1986 paper entitled ``Beta-delayed proton emission in the lanthanide region'' \cite{1986Wil01}. A 325 MeV $^{58}$Ni beam from the Berkeley Super HILAC bombarded a $^{92}$Mo target and $^{144}$Ho was produced in the fusion-evaporation reaction $^{92}$Mo($^{58}$Ni,3p2n). Beta-delayed particles, X-rays and $\gamma$-rays were measured following mass separation with the on-line isotope separator OASIS. ``A short lived proton emitter with a half-life of 0.7$\pm$0.1~s was assigned to the new isotope $^{144}$Ho on the basis of Dy K x-rays observed in coincidence with the protons.'' The quoted half-life is the currently accepted value.

\subsection*{$^{145}$Ho}\vspace{0.0cm}
Goettig et al.\ published the observation of $^{145}$Ho in the 1987 paper titled ``Decoupled bands in odd-A rare-earth nuclei below N=82'' \cite{1987Goe01}. Enriched $^{92}$Mo and $^{96}$Ru targets were bombarded with 250 MeV $^{56}$Fe and 240 MeV $^{52}$Cr beams, from the Daresbury Tandem Van de Graaff accelerator, respectively, and $^{145}$Ho was produced in the (p2n) fusion evaporation reaction. Neutrons, charged particles and $\gamma$-rays were measured with a 37 element wall, a Si surface barrier telescope and four Compton suppressed Ge detectors, respectively. ``In conclusion we have observed bands in the odd-proton nuclei $^{137}$Eu, $^{141,143}$Tb, $^{145}$Ho and the odd-neutron nucleus $^{145}$Dy, all of which are nuclei where no information on excited states was previously available.''

\subsection*{$^{146}$Ho}\vspace{0.0cm}
$^{146}$Ho was observed by Gui et al.\ and the results were published in the 1982 paper ``150~ms 10$^{+}$ isomer in $^{146}$Dy'' \cite{1982Gui01}. The Munich heavy-ion postaccelerator was used to bombard $^{90}$Zr and $^{91}$Zr targets with 233 and 250~MeV $^{58}$Ni beams. Gamma-ray spectra were measured with a coaxial and a planar Ge(Li) detector and conversion electrons were detected with a solenoidal spectrometer. ``The 3.9~s activity, observed for the identified $^{146}$Dy lines in the residual activities has been attributed to the $\beta$ decay of $^{146}$Ho.'' This value is in agreement with the currently accepted value of 3.6(3)~s.

\subsection*{$^{147}$Ho}\vspace{0.0cm}
``Very proton rich nuclei with N$\approx$82'' was published in 1982 by Nolte et al.\ documenting the observation of $^{147}$Ho \cite{1982Nol02}. $^{58}$Ni beams of energies of 233$-$250~MeV from the Munich MP tandem and heavy-ion linear rf post accelerator were used to bombard $^{92}$Mo targets forming $^{147}$Ho in the fusion-evaporation reaction $^{92}$Mo($^{58}$Ni,3p). Gamma-ray singles and coincidences were measured with coaxial and planar Ge(Li) detectors. ``A half-life of 5.8$\pm$0.4~s was obtained for the new isotope $^{147}$Ho.'' This half-life is the currently accepted value.

\subsection*{$^{148,149}$Ho}\vspace{0.0cm}
In the 1979 paper ``Identification of $^{148}$Ho and $^{149}$Ho'' Toth et al.\ reported the discovery of $^{148}$Ho and $^{149}$Ho \cite{1979Tot01}. A $^{10}$B beam was accelerated up to 101 MeV by the Texas A\&M isochronous cyclotron and bombarded an enriched $^{144}$Sm target. A helium gas-jet system transported the reaction products to a counting station where $\gamma$-ray singles and coincidences were recorded with large-volume Ge(Li) detectors. ``The 9-s isotope, $^{148}$Ho, was identified mainly through a 1688-keV $\gamma$-ray which: (1) was in coincidence with dysprosium K x rays, (2) increased dramatically in intensity when the $^{10}$B bombarding energy was raised from 85 to 96~MeV, and (3) remained constant (over the same range) in intensity relative to that of the 620-keV $\gamma$ ray known to belong to $^{148}$Dy decay.'' This half-life corresponds to an isomeric state. ``The 21-s isotope, $^{149}$Ho, was found to decay primarily to a $^{149}$Dy level at 1091~keV and less intensely to the i$_{13/2}$ 1073-keV state observed in a previous in-beam $\gamma$-ray study.'' This half-life is consistent with the currently accepted value of 21.1(2)~s.

\subsection*{$^{150-153}$Ho}\vspace{0.0cm}
Macfarlane and Griffioen identified $^{150}$Ho, $^{151}$Ho, $^{152}$Ho, and $^{153}$Ho as reported in the 1963 paper ``Alpha decay properties of some holmium isotopes near the 82-neutron closed shell'' \cite{1963Mac02}. $^{16}$O beams were accelerated to 75--137 MeV by the Berkeley HILAC and bombarded $^{141}$Pr targets to produce holmium isotopes in the fusion-evaporation reactions $^{141}$Pr($^{16}$O,xn). A Frisch-grid ionization chamber was used to measured subsequent $\alpha$ decays. ``Some Dy$^{150}$ alpha activity was also observed on the plates, which is undoubtedly the result of Ho$^{150}$ decay. From the level of Dy$^{150}$ activity on the two plates, Ho$^{150}$ appears to have a half-life of approximately 20 sec... This indicates that the parent, Ho$^{151}$, has a half-life of approximately 30 sec which is consistent with the value obtained for the 4.51-MeV group... This means that the parent Ho$^{l52}$ has a half-life of approximately 1 min, a value which is consistent with the measured half-life of the 4.45-MeV alpha group... A search was made for evidence of Ho$^{153}$ alpha activity by looking at products of the Pr$^{14l}$+0$^{16}$ reaction at incident energies in the range of 70 to 90 MeV. At these low energies a small peak was observed at an alpha particle energy of 3.92 MeV which was found to decay with a half-life of 9 min.'' The $\sim$20~s and 9(2)~min half-lives reported for $^{150}$Ho and $^{153}$Ho correspond to isomeric states. For $^{151}$Ho and $^{152}$Ho Macfarlane and Griffioen identified the ground states as well as isomeric states.

\subsection*{$^{154}$Ho}\vspace{0.0cm}
The first observation of $^{154}$Ho was reported by Lagarde et al.\ in the 1966 paper ``D\'esint\'egration de quelques isotopes d'erbium et d'holmium d\'eficients en neutrons'' \cite{1966Lag01}. Natural holmium oxide targets were irradiated with protons from the Orsay synchrocyclotron producing erbium isotopes in (p,xn) reactions. Reaction products were isotopically separated with a double-deflection magnetic separator. $^{154}$Ho was then populated in the decay of $^{154}$Er. Gamma-rays were measured with NaI(Tl) and Ge(Li) detectors and conversion electrons were measured with a silicon surface barrier detector. ``La p\'eriode de $^{154}$Ho est \'egale \`a 7$\pm$1 mn et le spectre $\gamma$ manifeste un pic vers 350 keV.'' [The period of $^{154}$Ho is equal to 7$\pm$1 mn and the $\gamma$ spectrum exhibits a peak around 350 keV.] This half-life is close to the currently accepted value of 11.76(19)~min.

\subsection*{$^{155}$Ho}\vspace{0.0cm}
In 1959, the discovery of $^{155}$Ho was announced by Kalyamin et al.\ in ``New positron activities in neutron-deficient isotopes of lutetium, ytterbium, and holmium'' \cite{1959Kal01}. A tantalum target was irradiated by 660-Mev protons from the Dubna synchrocyclotron. Decay curves and positron spectra were measured following chemical separation. ``In the daughter fraction of dysprosium (chromatographically separated from the holmium fraction), the characteristic $\gamma$ spectrum of Dy$^{155}$ was found, using a scintillation $\gamma$ spectrometer.  This makes more plausible the supposition that the mass number of the new isotope of holmium is 155.'' The quoted half-life of $\sim$50~min agrees with the currently accepted value of 48(1)~min.

\subsection*{$^{156}$Ho}\vspace{0.0cm}
``Nuclear spectroscopy of neutron-deficient rare earths (Tb through Hf)'' was published in 1957 by Mihelich et al.\ describing the observation of $^{156}$Ho \cite{1957Mih01}. A dysprosium oxide target was irradiated with 22 MeV protons from the ORNL 86-inch cyclotron. The resulting activities were measured with a conversion electron spectrograph and a scintillation counter following chemical separation. ``Ho$^{156}$($\sim$1~hr)$\rightarrow$Dy$^{156}$.---As mentioned before, there is evidence for a Ho$^{156}$ activity of 1-hr half-life which decays by electron capture to a 138-kev level in Dy$^{156}$.'' This half-life is in agreement with the currently accepted value of 56(1)~min.

\subsection*{$^{157}$Ho}\vspace{0.0cm}
In the 1966 paper ``New isotopes of Er$^{157}$, Ho$^{157}$, and Er$^{156}$'' Zhelev et al.\ announced the discovery of $^{157}$Ho \cite{1966Zhe01}. 660-MeV protons from the Dubna synchrocyclotron irradiated a tantalum target. Gamma spectra were measured with a scintillation spectrometer following chemical separation. ``We determined the Ho$^{157}$ half-life from the amount of Dy$^{157}$ activity accumulated from Ho$^{157}$ in three successive separations. Our result was 18$^{+2}_{-4}$~min.'' This result reasonably agrees with the accepted value 12.6(2)~min. Just 2~months later Lagarde and Gizon independently reported their discovery of $^{157}$Ho with a half-life of 14(1)~min \cite{1966Lag01}.

\subsection*{$^{158}$Ho}\vspace{0.0cm}
$^{158}$Ho was first observed by Dneprovskii in 1961 as reported in ``New isotopes of holmium and erbium'' \cite{1961Dne01}. Tantalum targets were bombarded with 660~MeV protons from the Dubna synchrocyclotron and $^{158}$Ho was populated by $\beta$-decay from $^{158}$Er produced in the reaction. Conversion electrons were measured with a magnetic beta-spectrograph following chemical separation. ``Upon isolation of the daughter holmium isotope from the erbium fraction 2~hr after the latter was isolated from tantalum, the intensity of lines corresponding to $\gamma$ transitions had dropped one-half within (27$\pm$2)~min. From the facts accumulated, we may ascertain the presence of the decay chain Er $\stackrel{2.4hr}{\longrightarrow}$ Ho $\stackrel{27min}{\longrightarrow}$ Dy.'' This half-life corresponds to an isomeric state.

\subsection*{$^{159}$Ho}\vspace{0.0cm}
``A new holmium activity, $^{159}$Ho'' was published in 1958 announcing the discovery of $^{159}$Ho by Toth \cite{1958Tot01}. The Berkeley 60-in.\ cyclotron was used to bombard a Tb$_2$O$_3$ target with 48 MeV  $\alpha$-particles forming $^{159}$Ho through an ($\alpha$,4n) reaction. Decay curves were measured with a Geiger counter and X- and $\gamma$-ray spectra were measured with a NaI(Tl) scintillation spectrometer. ``From the evidence presented, one can draw the following conclusion: an activity that had at least four characteristic $\gamma$-rays was seen in the holmium fraction at full bombarding energy. Each of the photopeaks decayed with a half-life of about 33 min. The four $\gamma$-transitions were missing in the holmium fraction when the experiment was carried out below the ($\alpha$,4n) threshold. The new activity must therefore be $^{159}$Ho.'' This half-life agrees with the currently accepted value of 33.05(11)~min.

\subsection*{$^{160}$Ho}\vspace{0.0cm}
Wilkinson and Hicks published the observation of $^{160}$Ho in the 1950 paper ``Radioactive isotopes of the rare earths. III. Terbium and holmium isotopes'' \cite{1950Wil03}. Terbium targets were bombarded with 38 MeV $\alpha$-particles from the Berkeley 60-in.\ cyclotron. Electrons, positrons, $\gamma$-rays and X-rays were measured following chemical separation. ``22.5$\pm$0.5-min Ho$^{160}$: This activity was observed only in short 38-Mev $\alpha$-particle bombardments of terbium together with the 4.6-hr. and 65-day activities.'' This half-life is close to the currently adopted values of 25.6(3)~min. Other holmium assignments reported by Wilkinson and Hicks were subsequently shown to be incorrect by Handley in 1954 \cite{1954Han04}, however, Handley confirmed the assignment of $^{160}$Ho \cite{1954Han04}.

\subsection*{$^{161}$Ho}\vspace{0.0cm}
$^{161}$Ho was discovered in 1954 by Handley and Olson in the paper ``New radioactive nuclides of the rare earths'' \cite{1954Han02}. Erbium oxide was bombarded with 24-MeV protons from the Oak Ridge 86-in. cyclotron. Decay curves and $\gamma$-ray spectra were measured following chemical separation. ``Decay of the holmium fraction of the second separation was followed, and a half-life of 2.5 hours was observed... This activity is assigned a mass of 161 because it is the daughter of 3.6-hour Er$^{161}$.'' This half-life is in agreement with the currently accepted value 2.48(5)~h. A previously reported half-life of 4.6(1)~h \cite{1950Wil03} was evidently incorrect \cite{1954Han04}.

\subsection*{$^{162}$Ho}\vspace{0.0cm}
``Nuclear spectroscopy of neutron-deficient rare earths (Tb through Hf)'' was published in 1957 by Mihelich et al.\ describing the observation of $^{162}$Ho \cite{1957Mih01}. A dysprosium oxide target was irradiated with 22 MeV protons from the ORNL 86-inch cyclotron. The resulting activities were measured with a conversion electron spectrograph and a scintillation counter following chemical separation. ``Ho$^{162}$(67~min)$\rightarrow$Dy$^{162}$: We have listed the transitions assigned to Ho$^{162}$ in [the table]. The mass assignment is made on the basis of yields from targets enriched in various masses.'' This half-life agrees with the currently adopted value of 67.0(7)~min for an isomeric state.  A previously reported half-life of 65.0(5)~d \cite{1950Wil03} was evidently incorrect \cite{1954Han04}.

\subsection*{$^{163}$Ho}\vspace{0.0cm}
Hammer and Stewart published the observation of $^{163}$Ho in the 1957 paper ``Isomeric transitions in the rare-earth elements'' \cite{1957Ham01}. Holmium oxides were irradiated with x-rays from the 75 MeV Iowa State College synchrotron and $^{163}$Ho was produced in the photonuclear ($\gamma$,2n) reaction. Decay curves and X- and $\gamma$-ray spectra were recorded. ``Since Ho is a single isotope of mass 156, the ($\gamma$,2n) reaction would place the isomeric state in Ho$^{163}$ and the ($\gamma$,d) reaction would pace it at Dy$^{163}$. However, Dy$^{l64}$ is a stable isotope and if the 0.8-sec isomeric state were in Dy$^{163}$, we should have observed it from the Dy$^{164}$($\gamma$,n)Dy$^{163}$ reaction. Since we did not, we assume that the reaction we observed was Ho$^{166}$($\gamma$,2n)Ho$^{163m}$.''  The observed half-life of 0.8(1)~s is close to the currently adopted value of 1.09(3)~s for an isomeric state. A previously reported half-life of 5.20(5)~d \cite{1950Wil03} was evidently incorrect \cite{1954Han04}.

\subsection*{$^{164}$Ho}\vspace{0.0cm}
The first detection of $^{164}$Ho was reported in 1938 by Pool and Quill in ``Radioactivity induced in the rare earth elements by fast neutrons'' \cite{1938Poo02}. Fast neutrons produced with 6.3 MeV deuterons from the University of Michigan cyclotron irradiated holmium-rich yttrium and pure yttrium targets for comparison. Decay curves were measured with a Wulf string electrometer. ``The 47-min.\ period emits electrons and is assigned to Ho$^{164}$.'' This half-life is within a factor of two of the currently accepted values of 29(1)~min for the ground state and 38.0(10)~min for an isomeric state.

\subsection*{$^{165}$Ho}\vspace{0.0cm}
In 1934, Aston reported the first observation of stable $^{165}$Ho in ``The isotopic constitution and atomic weights of the rare earth elements'' \cite{1934Ast04}. Rare earth elements were analyzed with the Cavendish mass spectrograph. ``Holmium (67) is quite definitely simple 165.''

\subsection*{$^{166}$Ho}\vspace{0.0cm}
$^{166}$Ho was identified in 1936 by Hevesy and Levi reported in the paper ``The action of neutrons on the rare earth elements'' \cite{1936Hev01}. Holmium was irradiated by a 200-300 mCi radon-beryllium source and the activity was measured following chemical separation. ``Holmium has one stable isotope, 165; the activity observed is therefore due to the decay of $^{166}_{ 67}$Ho, the intensity of the activity observed being 20 per cent of that of dysprosium.'' The reported half-life of 35~h is within a factor of two of the currently accepted value 26.824(12)~h. Previously, the 35~h \cite{1935Hev01} and a 33~h \cite{1935Neu01} were reported without mass assignment. The report of a 2.6(2)~h half-life assigned to $^{166}$Ho \cite{1935Mar01} was evidently incorrect \cite{1936Hev01}.

\subsection*{$^{167}$Ho}\vspace{0.0cm}
In 1955 Handley et al.\ described the discovery of $^{167}$Ho in the paper ``Holmium-167'' \cite{1955Han01}. Er$_2$O$_3$ targets were bombarded with 22.4 MeV protons from the Oak Ridge 86-in.\ cyclotron. Decay curves, absorption spectra, and $\gamma$-ray spectra were measured with an end-window GM tube following chemical separation. ``From the differences in isotopic abundance it was calculated that the 3-hour activity was produced from Er$^{167}$, which would again assign to it a mass of 167. Thus, with the chemistry identifying it as holmium, it is given an assignment of Ho$^{167}$.'' The quoted half-life agrees with the currently accepted value 3.1(1)~h.

\subsection*{$^{168}$Ho}\vspace{0.0cm}
$^{168}$Ho was first observed by Wille and Fink in 1960 as reported in ``Activation cross sections for 14.8-Mev neutrons and some new radioactive nuclides in the rare earth region'' \cite{1960Wil02}. An erbium metal target enriched in $^{168}$Er was irradiated with neutrons produced in the $^{3}$H(d,n)$^{4}$He reaction from the University of Arkansas Cockcroft-Walton accelerator. Decay curves were measured with an aluminum-walled, methane-flow beta-proportional counter and $\gamma$-spectra were measured with a Na(Tl) detector. ``Irradiations of 98\% pure natural erbium metal exhibited a new 3.3$\pm$0.5-min half-life which could not be immediately assigned... For these reasons, we assign Ho$^{168}$ to the new 3.3-min activity.''  This half-life is used in the calculation of the currently accepted value of 2.99(7)~min.

\subsection*{$^{169}$Ho}\vspace{0.0cm}
Miyano and Koryanagi reported the discovery of $^{169}$Ho in the 1963 paper ``The new nucleide holmium-169'' \cite{1963Miy01}. Erbium oxide targets enriched in $^{170}$Er were irradiated with bremsstrahlungs $\gamma$-rays produced with the JAERI electron linear accelerator and $^{169}$Ho was produced by the ($\gamma$,p) reactions. Gamma- and beta-ray spectra were measured with a scintillation spectrometer and an anthracene crystal, respectively. ``Holmium-169 was produced from the $^{170}$Er($\gamma$,p) reaction with 21 MeV bremsstrahlung... The half life of this new nuclide was determined to be T$_{1/2}$ = 4.8$\pm$0.1~min.'' This half-life is included in the calculation of the currently accepted value of 4.72(10)~min. A previously reported half-life of 44~min \cite{1950But01} was evidently incorrect.

\subsection*{$^{170}$Ho}\vspace{0.0cm}
$^{170}$Ho was first observed by Wille and Fink in 1960 as reported in ``Activation cross sections for 14.8-Mev neutrons and some new radioactive nuclides in the rare earth region'' \cite{1960Wil02}. An erbium oxide target enriched in $^{170}$Er was irradiated with neutrons produced in the $^{3}$H(d,n)$^{4}$He reaction from the University of Arkansas Cockcroft-Walton accelerator. Decay curves were measured with an aluminum-walled, methane-flow beta-proportional counter. ``Therefore, it seems likely that the 40-sec activity is Ho$^{170}$ from the Er$^{170}$(n,p) reaction and the 4.4-min period is Dy$^{167}$ from the Er$^{170}$(n,$\alpha$) reaction.''  The quoted half-life of 40(10)~s is in agreement with the currently accepted value 43(2)~s for an isomeric state.

\subsection*{$^{171}$Ho}\vspace{0.0cm}
In 1988 the first observation of $^{171}$Ho was reported in ``Identification of the neutron-rich isotope $^{174}$Er'' by Chasteler et al.\ \cite{1989Cha01}. Natural tungsten targets were bombarded with a 8.5 MeV/u $^{176}$Yb beam from the Berkeley SuperHILAC and $^{171}$Ho was produced in multi-nucleon transfer reactions. X-, $\beta$- and $\gamma$-rays were measured with silicon, plastic scintillator, and germanium detector at the OASIS mass separation facility. ``Based on our measurements of the Er x rays in coincidence with the $\beta^{-}$ particles and the $^{171}$Er transitions, we can unambiguously assign the 49(5)~s to the $\beta^{-}$ decay of $^{171}$Ho and eliminate the $^{171m}$Er alternative interpretation.'' This half-life agrees with the currently accepted value 53(2)~s. A month later Rykaczewski et al.\ assigned a 47(5)~s half-life to either $^{171}$Ho or $^{171m}$Er \cite{1989Ryk01}.

\subsection*{$^{172}$Ho}\vspace{0.0cm}
In their 1991 paper, ``Investigation of the new isotope $^{172}$Ho and of $^{174}$Er'', Becker et al.\ announced the discovery of $^{172}$Ho \cite{1991Bec01}. A 11.6~MeV/u $^{136}$Xe from the GSI UNILAC accelerator bombarded a target of tungsten foils enriched in $^{186}$W, and $^{172}$Ho was produced in multinucleon transfer reactions. X-, $\beta$- and $\gamma$-ray spectra were measured at the GSI on-line mass separator. ``For $^{172}$Ho, the half-life was measured to be 25(3)~s and a decay scheme is proposed on the basis of $\gamma \gamma$-coincidence data.'' The quoted half-life is the currently accepted value.


\section{Discovery of $^{144-175}$Er}
Thirty-two erbium isotopes from A = 144$-$175 have been discovered so far; these include 6 stable, 20 neutron-deficient and 6 neutron-rich isotopes. According to the HFB-14 model \cite{2007Gor01}, on the neutron-rich side the bound isotopes should reach at least up to $^{224}$Er and on the neutron deficient side five more isotopes ($^{140-144}$Er) should be particle bound. In addition, seven more isotopes ($^{133-139}$Er) could still have half-lives longer than 10$^{-9}$~s \cite{2004Tho01}. Thus, about 61 isotopes have yet to be discovered corresponding to 66\% of all possible erbium isotopes.

\begin{figure}
	\centering
	\includegraphics[scale=.7]{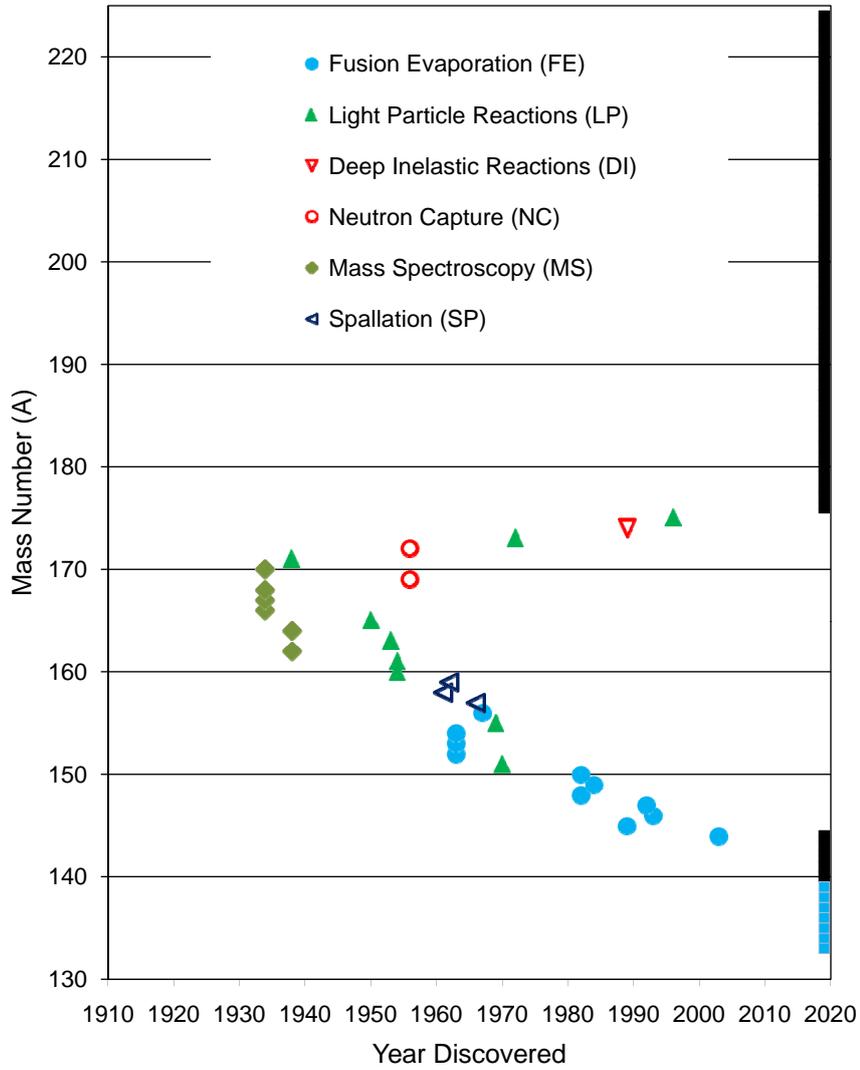}
	\caption{Erbium isotopes as a function of time when they were discovered. The different production methods are indicated. The solid black squares on the right hand side of the plot are isotopes predicted to be bound by the HFB-14 model. On the proton-rich side the light blue squares correspond to unbound isotopes predicted to have half-lives larger than $\sim 10^{-9}$~s.}
\label{f:year-er}
\end{figure}

Figure \ref{f:year-er} summarizes the year of discovery for all erbium isotopes identified by the method of discovery. The range of isotopes predicted to exist is indicated on the right side of the figure. The radioactive erbium isotopes were produced using fusion evaporation reactions (FE), light-particle reactions (LP), deep-inelastic reactions (DI), neutron capture (NC), and spallation (SP). The stable isotopes were identified using mass spectroscopy (MS). The discovery of each erbium isotope is discussed in detail and a summary is presented in Table 1.

\subsection*{$^{144}$Er}\vspace{0.0cm}
In the 2003 paper ``Fine structure in proton emission from $^{145}$Tm discovered with digital signal processing'', first evidence of $^{144}$Er was reported by Karny et al.\ \cite{2003Kar02}. A 315 MeV $^{58}$Ni beam bombarded a $^{92}$Mo target forming $^{145}$Tm which populated $^{144}$Er by proton emission. Recoil products were separated with the Oak Ridge Recoil Mass Separator RMS and implanted in a double-sided silicon strip detector which also recorded the proton decays. ``Since the daughter activity is an even-even nucleus, $^{144}$Er, the interpretation of the 1.73 and 1.40 MeV lines as transitions to the I$^\pi$ = 0$^+$ ground state and to the previously unknown I$^\pi$ = 2$^+$ excited state at 0.33(1)~MeV is obvious.'' An earlier report of the observation of $^{144}$Er was only published in a conference proceeding \cite{2000Sou01}.

\subsection*{$^{145}$Er}\vspace{0.0cm}
``Identification of $^{145}$Er and $^{145}$Ho'' was published in 1989 by Vierinen et al.\ describing the observation of $^{145}$Er \cite{1989Vie01}. Enriched $^{92}$Mo targets were bombarded with a 283-MeV $^{58}$Ni beam from the Berkeley SuperHILAC and $^{145}$Er was formed in the (2p3n) fusion-evaporation reaction. Recoil products were separated with the On-line Aparatur for SuperHILAC Isotope Separation OASIS and $^{145}$Er was identified by measuring $\gamma$-ray and delayed proton spectra. ``With the $^{145}$Dy half-life fixed at 8~s and assuming a negligible contribution from the potential $^{145}$Ho protons, a two-component analysis of the decay curves associated with the delayed protons in the 1.6- and 4-s tape cycles yielded a half-life of 0.9$\pm$0.3~s for $^{145}$Er.'' The quoted half-life is the currently accepted value for the isomeric state.

\subsection*{$^{146}$Er}\vspace{0.0cm}
$^{146}$Er was first observed by Toth et al.\ in 1993 as reported in ``Observation of $^{146}$Er electron capture and $\beta ^{+}$ decay'' \cite{1993Tot01}. A 280 MeV $^{58}$Ni beam from the Berkeley SuperHILAC bombarded a $^{92}$Mo target and $^{146}$Er was formed in the (2p2n) fusion-evaporation reaction. Recoil products were separated with the on-line facility OASIS. Coincidences between $\beta$-delayed protons, $\gamma$ rays, and x rays were recorded. ``From the time distribution of the x-ray events seen in these total-projected spectra the half-life of $^{146}$Er was determined to be 1.7(6)~s.'' This half-life is the currently accepted value.

\subsection*{$^{147}$Er}\vspace{0.0cm}
In the 1992 article ``Identification of the N=79 $^{147}$Er nucleus through $\gamma$-recoil coincidences'' de Angelis et al.\ reported the discovery of $^{147}$Er \cite{1992deA01}. A $^{92}$Mo target was bombarded with a 260 MeV $^{58}$Ni beam and $^{147}$Er was formed in the (2pn) fusion-evaporation reaction. Recoil products were separated with the Legnaro Recoil Mass Spectrometer RMS and identified with $\gamma$-recoil measurements. ``We have identified for the first time $\gamma$-rays in $^{147}$Er by means of $\gamma$-recoil coincidences at the Recoil Mass Spectrometer (RMS) of the Laboratori Nazionali di Legnaro (LNL).''

\subsection*{$^{148}$Er}\vspace{0.0cm}
``Very proton rich nuclei with N$\approx$82'' was published in 1982 by Nolte et al.\ documenting the observation of $^{148}$Er \cite{1982Nol02}. $^{58}$Ni beams of energies between 233 and 250~MeV from the Munich MP tandem and heavy-ion linear rf post accelerator were used to bombard $^{92}$Mo targets forming $^{148}$Er in the fusion-evaporation reaction $^{92}$Mo($^{58}$Ni,2p). Gamma-ray singles and coincidences were measured with coaxial and planar Ge(Li) detectors. ``With the help of cross bombardments and excitation functions, the two $\gamma$ lines at 244.2 and 315.2 keV were tentatively assigned to the $\beta$ decay of $^{148}$Er to $^{148}$Ho. The decay curve of the 244.2 keV line is also drawn in [the figure]. The corresponding half-life was found to be 4.5$\pm$0.4~s.'' This half-life is used in the calculation of the currently accepted value, 4.6(2)~s.

\subsection*{$^{149}$Er}\vspace{0.0cm}
$^{149}$Er was observed by Toth et al.\ and the results were published in the 1984 paper ``Beta-delayed proton activities: $^{147}$Dy and $^{149}$Er'' \cite{1984Tot01}. A 162 MeV $^{12}$C beam from the Berkeley 88-in.\ cyclotron bombarded a samarium oxide target enriched in $^{144}$Sm and $^{149}$Er was produced in the (7n) fusion-evaporation reaction. A helium gas-jet apparatus was used to transport the recoil products to a collection box where $\gamma$-rays, X-rays and delayed protons were measured. ``Therefore, we assign the 9-sec activity to the $\beta$ decay of the hitherto unidentified isotope $^{149}$Er. Further, we assume that the 9-sec half-life is due primarily (though not exclusively) to the $h_{11/2}$ isomer rather than the $s_{1/2}$ ground state, since the high-spin species should be the predominant product in a heavy-ion induced compound nuclear reaction.'' This value is in agreement with the currently accepted value of 8.9(2)~s for the isomeric state.

\subsection*{$^{150}$Er}\vspace{0.0cm}
``Very proton rich nuclei with N$\approx$82'' was published in 1982 by Nolte et al.\ documenting the observation of $^{150}$Er \cite{1982Nol02}. $^{58}$Ni beams of energies between 233 and 250~MeV from the Munich MP tandem and heavy-ion linear rf post accelerator were used to bombard $^{94}$Mo targets forming $^{150}$Tm in the fusion-evaporation reaction $^{94}$Mo($^{58}$Ni,np). $^{150}$Er was then populated by the $\beta$ decay of $^{150}$Tm. Gamma-ray singles and coincidences were measured with coaxial and planar Ge(Li) detectors. ``With the help of measured excitation functions, cross bombardments and coincidences with x-rays, the 475.8 keV line was assigned to the $\beta$ decay of $^{150}$Er to $^{150}$Ho. [The figure] shows the decay of this line after 30 s irradiations. The half-life of $^{150}$Er was found to be 18.5$\pm$0.7~s.'' This half-life is the currently accepted value. Within a few months after the submission by Nolte et al.\ (Feb. 11), half-lives of 20(2)~s (Moltz et al., May 14 \cite{1982Mol01}), 22(2)~s (Toth et al., July 14 \cite{1982Tot01}), and 17(2)~s (Batist et al.\ published in Nov. \cite{1982Bat01}), as well as a $\gamma$-ray spectrum  (Helppi et al.\, Mar. 5 \cite{1982Hel01}) were reported independently. Helppi et al.\ and Batist et al.\ were aware of the work by Nolte et al, referring to a conference proceeding and a preprint, respectively. Previously, the observation of two isomers were assigned to either $^{150}$Ho or $^{150}$Er \cite{1979Hag01}.

\subsection*{$^{151}$Er}\vspace{0.0cm}
The discovery of $^{151}$Er was reported in ``Production of rare-earth $\alpha$ emitters with energetic $^{3}$He particles; new isotopes: $^{151}$Er, $^{156}$Yb, and $^{157}$Yb'' by Toth et al.\ in 1970 \cite{1970Tot01}. Dysprosium oxide targets enriched in $^{156}$Dy were bombarded with a 102.1MeV $^3$He beam from the Oak Ridge isochronous cyclotron ORIC and $^{151}$Er was produced in (8n) reactions. Recoils were transported to a Si(Au) detector with a helium gastransport system where $\alpha$-decay spectra were measured. ``Least-squares analyses of these data indicated that the 47-sec $^{151m}$Ho arises in part from the decay of a nuclide with a half-life of 23$\pm$2~sec. This new activity, based on the parent-daughter relationship, was assigned to $^{151}$Er.'' This half-life is included in the weighted average to attain the currently accepted value of 23.5(20)~s.

\subsection*{$^{152-154}$Er}\vspace{0.0cm}
In the 1963 paper ``Alpha-decay properties of some erbium isotopes near the 82-neutron closed shell'' Macfarlane and Griffioen announced the discovery of $^{152}$Er, $^{153}$Er, and $^{154}$Er \cite{1963Mac01}. $^{142}$Nd targets were bombarded with $^{16}$O beams of 75$-$151 MeV from the Berkeley heavy-ion accelerator Hilac, and $^{152}$Er, $^{153}$Er, and $^{154}$Er were formed in (6n), (5n), and (4n) fusion-evaporation reactions, respectively. Recoil products were collected on a charged plate which was placed in Frisch-grid ionization chamber to detect $\alpha$-particles. ``Er$^{152}$: ...A strong group was observed at 4.80-MeV alpha-particle energy which decays with a half-life of 10.7$\pm$0.5 sec... Er$^{153}$: The second prominent erbium alpha activity that was observed has an alpha-particle energy of 4.67 MeV and a half-life of 36$\pm$2 sec... Er$^{154}$: When the Nd$^{142}$ target was bombarded with O$^{16}$ ions at incident energies between 80 and 110 MeV, a weak alpha group was observed at 4.15-MeV alpha-particle energy which decays with a half-life of 4.5$\pm$1.0~min.'' These half-lives of 10.7(5)~s, 36(2)~s, and 4.5(10)~min agree with the presently accepted values of 10.3(10)~s, 37.1(2)~s, and 3.73(9)~min for $^{152}$Er, $^{153}$Er, and $^{154}$Er, respectively.

\subsection*{$^{155}$Er}\vspace{0.0cm}
Toth et al.\ observed $^{155}$Er and published the results in their 1969 paper ``New erbium isotope, $^{155}$Er \cite{1969Tot01}. Dysprosium oxide targets enriched in $^{156}$Dy were irradiated with 72.6~MeV $\alpha$ particles from the Oak Ridge isochronous cyclotron and $^{155}$Er was populated in the (5n) reaction. Recoil products were collected with a beryllium catcher which was rotated in front of a Si(Au) detector to record subsequent $\alpha$ emission. ``The 4.01-MeV $\alpha$ peak which decays with a 5.3-min half-life is assigned to $^{155}$Er on the basis of the following evidence: (a) The peak was not observed in 20--60-MeV proton bombardments of $^{156}$Dy... (b) The shape of the calculated $^{156}$Dy($\alpha$,5n) excitation function followed closely the experimental excitation function for the production of this 4.01-MeV peak. A mass number of 155 appears thus likely for this new $\alpha$ emitter.'' The quoted half-life is the currently accepted value.

\subsection*{$^{156}$Er}\vspace{0.0cm}
Ward et al.\ published the identification of $^{156}$Er in the 1967 paper ``Gamma rays following $^{40}$Ar-induced reactions'' \cite{1967War02}. Isotopically enriched $^{120}$Sn targets were bombarded with $^{40}$Ar beams and $^{156}$Er was populated in the fusion-evaporation reaction $^{120}$Sn($^{40}$Ar,4n). Gamma-ray spectra were studied using a lithium-drifted germanium counter. ``Gamma-ray spectra from the reactions $^{124,122,120}$Sn($^{40}$Ar,4n)$^{60,58,56}$Er are shown in [the figure].'' The first five transition of the rotational ground-state band were measured for $^{156}$Er. Previously, only upper limits of 10$-$12~min \cite{1966Zhe01} and $<$4~min \cite{1966Lag01} for the half-life of $^{156}$Er were reported.

\subsection*{$^{157}$Er}\vspace{0.0cm}
In the 1966 paper ``New isotopes of Er$^{157}$, Ho$^{157}$, and Er$^{156}$'' Zhelev et al.\ announced the discovery of  $^{157}$Er \cite{1966Zhe01}. 660-MeV protons from the Dubna synchrocyclotron irradiated a tantalum target. Gamma spectra were measured with a scintillation spectrometer following chemical separation. ``We determined the half-life of Er$^{157}$ from the amount of the daughter isotope Dy$^{157}$ that was accumulated in successive separations.  There was no need to make any assumption regarding the decay constant of Ho$^{157}$.  The half-life of Er$^{157}$ was 24$^{+2}_{-4}$~min.'' This half-life is close to the currently accepted value of 18.65(10)~min. Just 2~months later Lagarde and Gizon independently reported their discovery of $^{157}$Er with a half-life of 25~min \cite{1966Lag01}.

\subsection*{$^{158}$Er}\vspace{0.0cm}
$^{158}$Er was first observed by Dneprovskii in 1961 as reported in ``New isotopes of holmium and erbium'' \cite{1961Dne01}. Tantalum targets were bombarded with 660~MeV protons from the Dubna synchrocyclotron and $^{158}$Er was populated in spallation reactions. Conversion electrons were measured with a magnetic beta-spectrograph following chemical separation. ``From the facts accumulated, we may ascertain the presence of the decay chain Er $\stackrel{2.4hr}{\longrightarrow}$ Ho $\stackrel{27min}{\longrightarrow}$ Dy... The position of the first excitation level as a function of neutron number suggests that the mass number of the nuclei belonging to the above decay chain is A=158.'' This 2.4~h half-life is in reasonable agreement with the currently accepted value of 2.29(6)~h.

\subsection*{$^{159}$Er}\vspace{0.0cm}
Abdurazakov et al.\ reported the discovery of $^{159}$Er in their 1961 paper ``A new isotope Er$^{159}$'' \cite{1962Abd01}. 660-MeV protons from the Dubna synchrocyclotron bombarded a tantalum target. Conversion electrons were measured from successive exposures of photographic films in the $\beta$ spectrograph. ``A new erbium isotope of mass number 159 (T$_{1/2} \sim$1~hr) has been discovered.'' The quoted half-life is within a factor of two of the currently accepted value of 36(1)~min and Abdurazakov et al.\ also identified conversion lines of the daughter nucleus $^{159}$Ho.

\subsection*{$^{160}$Er}\vspace{0.0cm}
``Mass assignments by isotope separation'' was published in 1954 by Michel and Templeton documenting the observation of $^{160}$Er \cite{1954Mic01}. The Berkeley 184-inch cyclotron was used to bombard tantalum targets with 350 MeV protons. The resulting activities were measured with a G-M counter and a scintillation spectrograph. ``In addition, the following new isotopes formed in the spallation of tantalum with 350-Mev protons have been assigned: Tm$^{165}$ (29~hours), Er$^{160}$ (30~hours), Er$^{161}$ (3.5~hours).'' The quoted half-life for $^{160}$Er is in reasonable agreement with the currently accepted value of 28.58(91)~h.

\subsection*{$^{161}$Er}\vspace{0.0cm}
$^{161}$Er was discovered in 1954 by Handley and Olson in the paper ``New radioactive nuclides of the rare earths'' \cite{1954Han02}. Erbium oxide was bombarded with 24-MeV protons from the Oak Ridge 86-in.\ cyclotron. Decay curves and $\gamma$-ray spectra were measured following chemical separation. ``Therefore, the 3.6-hour species is produced chiefly by (p,2n) with Tm$^{161}$ species being short-lived and decaying immediately to Er$^{161}$.'' This half-life is close to the currently accepted value 3.21(3)~h. Less than four months later, Michel and Templeton independently reported a half-life of 3.5~h \cite{1954Mic01}.

\subsection*{$^{162}$Er}\vspace{0.0cm}
Dempster reported the discovery of $^{162}$Er in the 1938 paper ``The isotopic constitution of gadolinium, dysprosium, erbium and ytterbium'' \cite{1938Dem01}. An erbium oxide sample reduced with lanthanum was used for analysis in the Chicago mass spectrograph. ``Two new isotopes were also observed in erbium reduced with lanthanum at masses 164 and 162, the first on eleven photographs with exposures of ten seconds to twenty minutes and the second on four photographs with seven to twenty minutes' exposure. An example of the mass spectrum is given in [the figure]. The abundances were estimated as approximately 2 percent for the mass at 164 and 0.25 percent for the mass at 162.''

\subsection*{$^{163}$Er}\vspace{0.0cm}
The discovery of $^{163}$Er was reported in ``Erbium$^{163}$ and Thulium$^{165}$'' by Handley and Olson in 1953 \cite{1953Han01}. Holmium oxide was bombarded with 24 MeV protons from the ORNL 86-in. cyclotron and $^{163}$Er was produced in the $^{165}$Ho(p,3n) reaction. Decay curves and $\gamma$-ray spectra were measured with a scintillation spectrometer  following chemical separation. ``A 75-minute activity produced by bombarding Ho$^{165}$ with protons is assigned to Er$^{163}$.'' This half-life agrees with the currently accepted value of 75.0(4)~min.

\subsection*{$^{164}$Er}\vspace{0.0cm}
Dempster reported the discovery of $^{164}$Er in the 1938 paper ``The isotopic constitution of gadolinium, dysprosium, erbium and ytterbium'' \cite{1938Dem01}. An erbium oxide sample reduced with lanthanum was used for analysis in the Chicago mass spectrograph. ``Two new isotopes were also observed in erbium reduced with lanthanum at masses 164 and 162, the first on eleven photographs with exposures of ten seconds to twenty minutes and the second on four photographs with seven to twenty minutes' exposure. An example of the mass spectrum is given in [the figure]. The abundances were estimated as approximately 2 percent for the mass at 164 and 0.25 percent for the mass at 162.''

\subsection*{$^{165}$Er}\vspace{0.0cm}
``Radioactive$^{165}$Er'' by Butement reported the observation of $^{165}$Er in 1950 \cite{1950But02}. A holmium oxide target was bombarded with 10 MeV protons from the Harwell cyclotron and $^{165}$Er was produced in the $^{165}$Ho(p,n)$^{165}$Er charge exchange reaction. The subsequent decay curve was measured with a Geiger counter following chemical separation. ``The radioactivitiy of the erbium decayed entirely with a half-life of 10.0$\pm$0.1~hrs.'' This half-life is consistent with the currently accepted value of 10.36(4)~h. A previous assignment of a 1.1~min half-life \cite{1938Poo02} was evidently incorrect. A 12~h half-life had been measured without a mass assignment \cite{1935Hev01} or assigned to $^{169}$Er \cite{1938Poo02}.

\subsection*{$^{166-168}$Er}\vspace{0.0cm}
In 1934, Aston reported the first observation of $^{166}$Er, $^{167}$Er, and $^{168}$Er in ``The isotopic constitution and atomic weights of the rare earth elements'' \cite{1934Ast04}. Rare earth elements were analyzed with the Cavendish mass spectrograph. ``Erbium is not so complex as it was at first supposed to be. The early samples used were evidently contaminated. A pure sample gave three strong lines, 166, 167, 168 and a weak fourth 170.''

\subsection*{$^{169}$Er}\vspace{0.0cm}
The identification of $^{169}$Er was described by Bisi et al.\ in the 1956 paper ``An investigation of the first rotational level of $^{169}$Tm \cite{1956Bis01}. Erbium oxide was irradiated with slow neutrons in the Harwell reactor. Gamma- and beta-ray spectra were measured with a scintillation spectrometer and a $\beta$-ray spectrometer, respectively. ``The intensity of the $\beta$-rays was followed over 30 days. The half-life was found to be T$_{1/2}$=(9.0$\pm$0.2) d.'' This half-life agrees with the currently accepted value of 9.392(18)~d. A previous assignment of a 12~h half-life \cite{1938Poo02} was evidently incorrect.

\subsection*{$^{170}$Er}\vspace{0.0cm}
In 1934, Aston reported the first observation of $^{170}$Er in ``The isotopic constitution and atomic weights of the rare earth elements'' \cite{1934Ast04}. Rare earth elements were analyzed with the Cavendish mass spectrograph. ``Erbium is not so complex as it was at first supposed to be. The early samples used were evidently contaminated. A pure sample gave three strong lines, 166, 167, 168 and a weak fourth 170.''

\subsection*{$^{171}$Er}\vspace{0.0cm}
The first identification of $^{171}$Er was reported in 1938 by Pool and Quill in ``Radioactivity induced in the rare earth elements by fast neutrons'' \cite{1938Poo02}. Fast and slow neutrons produced with 6.3 MeV deuterons from the University of Michigan cyclotron irradiated erbium oxide targets. Decay curves were measured with a Wulf string electrometer. ``Since the 5.1-hr period is not present with fast neutron bombardment, it is assigned to Er$^{171}$.'' This half-life is within a factor of two of the currently accepted values of 7.516(2)~h.

\subsection*{$^{172}$Er}\vspace{0.0cm}
The discovery of $^{172}$Er was announced in 1956 by Nethaway et al.\ in the paper ``New isotopes: Er$^{172}$ and Tm$^{172}$'' \cite{1956Net01}. Erbium oxide was irradiated with neutrons from the Idaho Materials Testing Reactor and $^{172}$Er was formed by sequential neutron capture from $^{170}$Er. Decay curves were measured following chemical separation. ``A least-squares analysis gave a value of 49.8~hr for the Er$^{172}$ half-life. In consideration of the errors inherent in this milking technique, the probable error is set at $\pm$1~hr.'' This half-life is included in the weighted average used to obtain the currently accepted value, 49.3(3)~h.

\subsection*{$^{173}$Er}\vspace{0.0cm}
Pursiheimo et al.\ observed $^{173}$Er in 1972 and reported their results in ``The decay of 1.4 Min $^{173}$Er \cite{1972Pur01}. $^{176}$Yb$_2$O$_3$ and Yb$_2$O$_2$ targets were irradiated with 14$-$15 MeV neutrons from the Helsinki SAMES T 400 neutron generator and $^{173}$Er was produced in (n,$\alpha$) reactions. Gamma- and beta-ray spectra were measured with Ge(Li), NaI(Tl) and plastic scintillation detectors. ``A 1.4$\pm$0.1~min activity which is assigned to the $\beta ^-$-decay of $^{173}_{ 68}$Er has been produced with 14-15 MeV neutrons through the reaction $^{176}$Yb(n,$\alpha$)$^{173}$Er.'' This half-life is the currently accepted value. A previously reported half-life of 12.0(3)~min \cite{1968Iho01} was evidently incorrect.

\subsection*{$^{174}$Er}\vspace{0.0cm}
In 1988 the first observation of $^{174}$Er was reported in ``Identification of the neutron-rich isotope $^{174}$Er'' by Chasteler et al.\ \cite{1989Cha01}. Natural tungsten targets were bombarded with a 8.5 MeV/u $^{176}$Yb beam from the Berkeley SuperHILAC and $^{174}$Er was produced in multi-nucleon transfer reactions. X-, $\beta$- and $\gamma$-rays were measured with silicon, plastic scintillator, and germanium detector at the OASIS mass separation facility. ``Twelve $\gamma$ rays with energies, relative intensities, and coincidences listed in [the table] are assigned to the decay of the new isotope $^{174}$Er based on the following observations: the $\gamma$ rays decayed with a 3.3(2)~min.\ half-life which does not match any of the known half-lives in this isobaric chain.'' This half-life agrees with the currently accepted value 3.2(2)~min.

\subsection*{$^{175}$Er}\vspace{0.0cm}
``The $\gamma$-decay of a new neutron-rich nucleus $^{175}$Er'' was published in 1996 announcing the discovery of $^{175}$Er by Zhang et al.\ \cite{1996Zha01}. Ytterbium metal targets were irradiated with 14 MeV neutrons which were produced by bombarding  a Ti$^3$H target with deuterons from the Lanzhou 600-kV Cockcroft-Walton accelerator. Two coaxial HPGe and a HPGe planar detector were used to measure $\gamma$-and X-rays, respectively. ``A 1.2$\pm$0.3~min activity produced with 14-MeV neutrons using the reaction $^{176}$Yb(n,2p)$^{175}$Er has been assigned to $\beta^-$ decay of the unreported erbium isotope $^{175}$Er.'' This half-life is the currently accepted value.


\section{Discovery of $^{145-177}$Tm}
Thirty-three thulium isotopes from A = 145$-$177 have been discovered so far; these include 1 stable, 24 neutron-deficient and 8 neutron-rich isotopes. According to the HFB-14 model \cite{2007Gor01}, on the neutron-rich side $^{228}$Tm should be the last odd-odd particle stable neutron-rich nucleus while the odd-even particle stable neutron-rich nuclei should continue at least through $^{231}$Tm. On the neutron deficient side the dripline has been crossed with the observation of proton emission of $^{145-147}$Tm. Five additional isotopes ($^{140-144}$Tm) could still have half-lives longer than 10$^{-9}$~s \cite{2004Tho01}. Thus, about 58 isotopes have yet to be discovered corresponding to 64\% of all possible thulium isotopes.

\begin{figure}
	\centering
	\includegraphics[scale=.7]{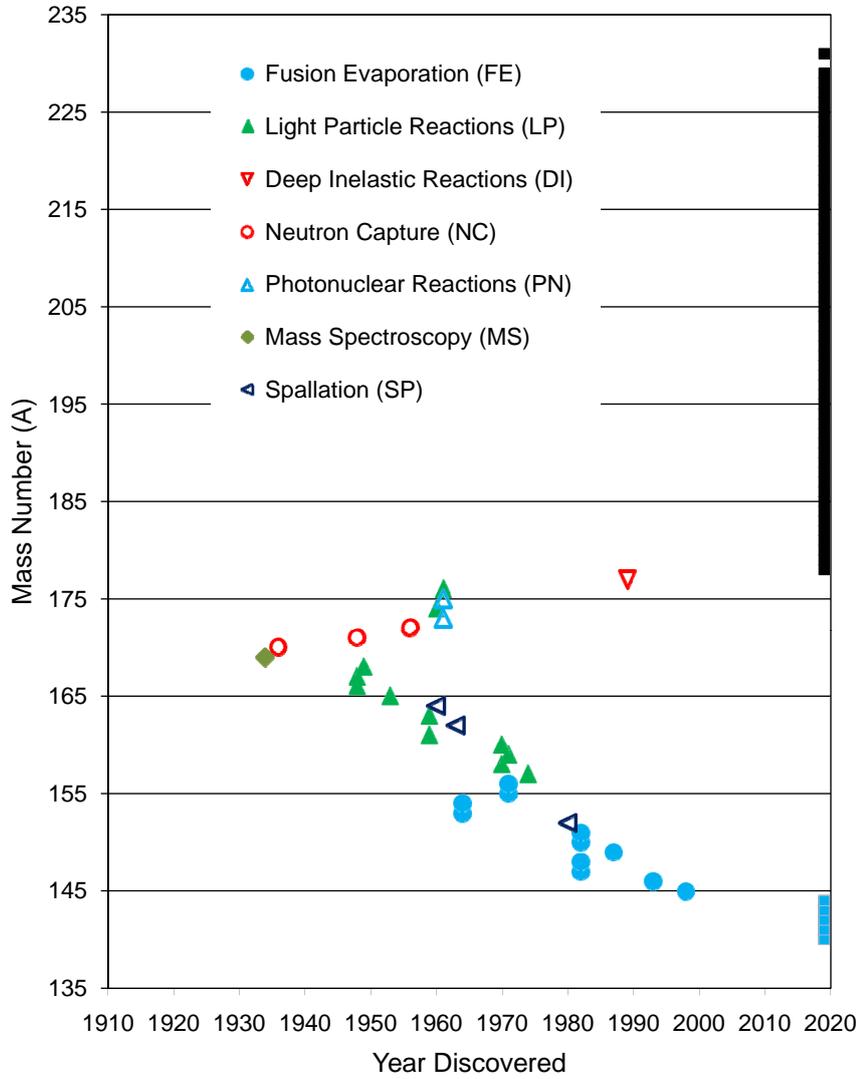}
	\caption{Thulium isotopes as a function of time when they were discovered. The different production methods are indicated. The solid black squares on the right hand side of the plot are isotopes predicted to be bound by the HFB-14 model. On the proton-rich side the light blue squares correspond to unbound isotopes predicted to have half-lives larger than $\sim 10^{-9}$~s.}
\label{f:year-tm}
\end{figure}

Figure \ref{f:year-tm} summarizes the year of discovery for all thulium isotopes identified by the method of discovery. The range of isotopes predicted to exist is indicated on the right side of the figure. The radioactive thulium isotopes were produced using fusion evaporation reactions (FE), light-particle reactions (LP), deep-inelastic reactions (DI), photo-nuclear reactions (PN), neutron capture (NC), and spallation (SP). The stable isotope was identified using mass spectroscopy (MS). The discovery of each thulium isotope is discussed in detail and a summary is presented in Table 1.

\subsection*{$^{145}$Tm}\vspace{0.0cm}
In the 1998 paper ``Observation of the exotic nucleus $^{145}$Tm via its direct proton decay'' Batchelder et al.\ announced the discovery of $^{145}$Tm \cite{1998Bat01}. A 315 MeV $^{58}$Ni beam from the Oak Ridge tandem accelerator bombarded an enriched $^{92}$Mo target and $^{145}$Tm was produced in the $^{92}$Mo($^{58}$Ni,p4n) fusion-evaporation reaction. Reaction products were separated with the Recoil Mass Spectrometer (RMS) and implanted in a double-sided silicon strip detector which also recorded subsequent proton emissions. ``In summary, we have observed direct proton emission from the 0$h_{11/2}$ ground state of $^{145}$Tm with E$_p$ and T$_{1/2}$ of 1.728(10)~MeV and 3.5(10)~$\mu$s respectively.'' This half-life agrees with the currently adopted value of 3.17(20)~$\mu$s.

\subsection*{$^{146}$Tm}\vspace{0.0cm}
``Proton radioactivity from $^{146}$Tm. The completion of a sequence of four odd-odd proton emitters'' announced the discovery of $^{146}$Tm  by Livingston et al.\ in 1993 \cite{1993Liv01}. An enriched $^{92}$Mo target was bombarded with a 287~MeV $^{58}$Ni beam and $^{146}$Tm was populated in the $^{92}$Mo($^{58}$Ni,p3n) fusion-evaporation reaction. Reaction products were separated with the Daresbury Recoil Separator and implanted in a double-sided silicon strip detector which also recorded subsequent proton emissions. ``On this basis, the two lines observed in the A=146 region are assigned to proton emission from $^{146}$Tm with corresponding Q-values of 1127$\pm$5~keV and 1197$\pm$5~keV. The measured half-lives of these lines are 235$\pm$27~ms and 72$\pm$23~ms, indicating that the lines originate from different states of $^{146}$Tm.'' Both of these half-lives correspond to currently accepted values, 235(27)~ms for the ground state, and 72(23)~ms for an isomeric state.

\subsection*{$^{147}$Tm}\vspace{0.0cm}
In 1982, Klepper et al.\ announced the discovery of $^{147}$Tm in ``Direct and beta-delayed proton decay of very neutron-deficient rare-earth isotopes produced in the reaction $^{58}$Ni+$^{92}$Mo'' \cite{1982Kle01}. A 4.6 MeV/u $^{58}$Ni beam from the GSI UNILAC bombarded a $^{92}$Mo target and $^{147}$Tm was formed in the $^{92}$Mo($^{58}$Ni,p2n) fusion-evaporation reaction. Reaction products were separated with the GSI on-line mass separator and implanted into carbon foils which were placed in front of surface barrier detector telescopes recording subsequent proton emissions.  ``Using the reaction $^{58}$Ni+$^{92}$Mo, a 1055~keV, 0.42~s proton activity was observed and preliminary assigned to $^{147}$Tm.'' This half-life of 0.42(10)~s is close to the currently accepted value of 0.58(3)~s.

\subsection*{$^{148}$Tm}\vspace{0.0cm}
``Very proton rich nuclei with N$\approx$82'' was published in 1982 by Nolte et al.\ documenting the observation of $^{148}$Tm \cite{1982Nol02}. $^{58}$Ni beams of energies between 233 and 250~MeV from the Munich MP tandem and heavy-ion linear rf post accelerator were used to bombard $^{92}$Mo targets forming $^{148}$Tm in the fusion-evaporation reaction $^{92}$Mo($^{58}$Ni,np). Gamma-ray singles and coincidences were measured with coaxial and planar Ge(Li) detectors. ``This activity was assigned to the new isotope $^{148}$Tm populated through the reaction $^{92}$Mo($^{58}$Ni,np). The $\gamma$ intensities of the observed lines are given in [the table]. The decay curve of the 646.6 keV line is displayed in [the figure]. A half-life of 0.7$\pm$0.2~s was obtained for $^{148}$Tm.'' This half-life is the currently accepted value.

\subsection*{$^{149}$Tm}\vspace{0.0cm}
``Observation of $^{149}$Tm decay to $^{149}$Er levels and $\beta$-delayed proton emission'' reported the discovery of $^{149}$Tm in 1987 by Toth et al.\ \cite{1987Tot02}. An enriched $^{94}$Mo target was bombarded with a 259 MeV $^{58}$Ni beam from the Berkeley SuperHILAC and $^{149}$Tm was formed in the fusion-evaporation reaction $^{94}$Mo($^{58}$Ni,p2n). Reaction products were separated with the OASIS on-line mass separator and $\gamma$-rays, X-rays and positrons were measured. ``A new activity (T$_{1/2}$=0.9$\pm$0.2~s), with at least seven $\gamma$ rays following its $\beta$ decay, was observed in the A=149 mass chain. It is assigned to the hitherto unknown isotope $^{149}$Tm because the $\gamma$ rays are in coincidence with Er K x rays and because several of them are also in coincidence with the 111.3-keV transition seen in $^{149}$Er$^{m}$ isomeric decay.'' This half-life is the currently accepted value.

\subsection*{$^{150}$Tm}\vspace{0.0cm}
``Very proton rich nuclei with N$\approx$82'' was published in 1982 by Nolte et al.\ documenting the observation of $^{150}$Tm \cite{1982Nol02}. $^{58}$Ni beams of energies between 233 and 250~MeV from the Munich MP tandem and heavy-ion linear rf post accelerator were used to bombard $^{94}$Mo targets forming $^{150}$Tm in the fusion-evaporation reaction $^{94}$Mo($^{58}$Ni,np). Gamma-ray singles and coincidences were measured with coaxial and planar Ge(Li) detectors. ``This activity was consequently assigned to the new isotope $^{150}$Tm. This isotope was produced through the reaction $^{94}$Mo($^{58}$Ni,np). The decay curve of the 207.5 keV line after 6~s irradiations is plotted in [the figure]. From this, a half-life of 3.5$\pm$0.6~s was obtained for $^{150}$Tm.'' This half-life is within a factor of two of the currently accepted value of 2.2(2)~s and the observed $\gamma$-transitions in the daughter nucleus $^{150}$Er were identified correctly.

\subsection*{$^{151}$Tm}\vspace{0.0cm}
In the 1982 paper ``Yrast $(\pi h_{11/2})^{n}$ excitations in proton rich N=82 nuclei'' Helppi et al.\ identified $^{151}$Tm \cite{1982Hel01}. $^{95}$Mo and $^{93}$Nb targets were bombarded with 255 MeV $^{58}$Ni and $^{60}$Ni beams from the Argonne Tandem-Linac forming $^{153}$Tm and $^{153}$Yb, respectively. $^{151}$Tm was then formed in (2n) and (1p1n) fusion-evaporation reactions. Gamma-ray spectra were measured with Ge(Li) and NaI detectors. ``Since in addition the lines were found to be coincident with Tm X-rays, they are assigned to the N=82 nucleus $^{151}$Tm. The detailed results showed that the four transitions occur in cascade, de-exciting an isomer with T$_{1/2}$=470$\pm$50~ns, and they established the E2 character for the 140~keV transition.''

\subsection*{$^{152}$Tm}\vspace{0.0cm}
The discovery of $^{152}$Tm was published in the 1980 paper ``New results in the decay of $^{150}$Ho and $^{152}$Tm'' by Liang et al.\ \cite{1980Lia01}. Metallic erbium targets were bombarded with 200 MeV protons from the Orsay synchrocyclotron. Reaction products were separated with the on-line isotope separator ISOCELE II and positrons and $\gamma$-ray spectra were measured. ``$^{152}$Tm: The analysis of the $\gamma$ spectra shows four transitions with energies of 279.9, 422.5, 672.6 and 808.2 keV decaying with the half-life of T$_{1/2}$=5.2$\pm$0.6~s.'' This half-life is the currently accepted value of an isomeric state of $^{152}$Tm.

\subsection*{$^{153,154}$Tm}\vspace{0.0cm}
``Alpha-decay properties of some thulium and ytterbium isotopes near the 82-neutron shell'' by Macfarlane announced the discovery of $^{153}$Tm and $^{154}$Tm in 1964 \cite{1964Mac01}. Praseodymium oxide and neodymium oxide (enriched in $^{142}$Nd) were bombarded with 131$-$195 MeV $^{20}$Ne and 121$-$185 MeV $^{19}$F beams from the Berkeley heavy-ion linear accelerator Hilac, respectively. $^{153}$Tm and $^{154}$Tm were formed in (8n) and (7n) fusion evaporation reactions and identified by measuring excitation functions and $\alpha$-decay spectra. ``The highest energy Tm alpha group observed has an alpha-particle energy of 5.11$\pm$0.02~MeV and decays with a half-life of 1.58~sec. This activity is tentatively assigned to the 84-neutron isotope Tm$^{153}$ on the basis of alpha-decay systematics... A second Tm alpha group decaying with a half-life of 2.98 sec was observed at an alpha-particle energy of 5.04 MeV. On the basis of alpha-decay systematics the most likely mass assignment appeared to be Tm$^{154}$.'' These half-lives 1.58(15)~s and 2.98(20)~s agree with the currently adopted values of 1.48(1)~s and 3.30(7)~s for the ground state of $^{153}$Tm and an isomeric state of $^{154}$Tm, respectively.

\subsection*{$^{155,156}$Tm}\vspace{0.0cm}
$^{155}$Tm and $^{156}$Tm were first observed by Toth et al.\ as reported in the 1971 paper ``Investigation of thulium $\alpha$ emitters; new isotopes $^{155}$Tm and $^{156}$Tm'' \cite{1971Tot01}. The Oak Ridge isochronous cyclotron was used to bombard enriched $^{144}$Sm and $^{147}$Sm targets with $^{14}$N beams of up to 103 MeV and $^{155}$Tm and $^{156}$Tm were formed in (3n) and (5n) fusion evaporation reactions, respectively. A helium gas system transported recoil products in front of a Si(Au) detector where subsequent $\alpha$-emission was detected. ``The sum of the two curves, labeled A$_1$(t)+A$_2$(t), is seen to agree with the data points; thus the decay data obtained at 65 MeV for the 4.60-MeV peak are certainly consistent with our assignment of the 4.45-MeV $\alpha$-particle group to $^{155}$Tm... Least squares analysis indicated a genetic relationship between two radioactive components, one with a  half-life of 2.38~min and the other with an 80-sec half-life. Because the 2.38-min half-life is that of one of the isomers of $^{152}$Ho, the parent-daughter relationship establishes the existence of a new thulium nuclide, $^{156}$Tm.'' The measured half-life of 39(3)~s for $^{155}$Tm was subsequently questioned \cite{1977Agu01}, however, later Toth et al.\ demonstrated that their first experiment was likely a sum of the ground state (21.6(2)~s) and an isomeric state (45(3)~s). The quoted half-life of 80(3)~s for $^{156}$Tm is used in the weighted average to obtain the currently accepted value 83.8(18)~s.

\subsection*{$^{157}$Tm}\vspace{0.0cm}
Putaux et al.\ reported the observation of $^{157}$Tm in the 1974 paper ``On-line separation of thulium isotopes'' \cite{1974Put01}. Erbium targets were irradiated with 157~MeV protons from the Orsay synchrocyclotron and $^{157}$Tm was produced in (p,xn) reactions. Residues were separated with the on-line ISOCELE separator and $\gamma$-ray spectra were measured with a Ge(Li) detector. ``Some short-life gamma radiations are attributed to $^{157}$Tm. We measured the half-life of $^{157}$Tm with two well separated $\gamma$-rays (110.3~keV and 241.6~keV) and we propose the half-life (200$\pm$25)~s.'' The quoted half-life is included in the calculation of the currently adopted value of 3.63(9)~min.

\subsection*{$^{158}$Tm}\vspace{0.0cm}
``New isotopes $^{158}$Tm and $^{160}$Tm'' was published in 1970 documenting the observation of $^{158}$Tm by de Boer et al.\ \cite{1970deB01}. Erbium oxide samples enriched in $^{162}$Er were irradiated with 54 MeV protons from the Amsterdam synchrocyclotron and $^{158}$Tm was populated in (p,5n) reactions. Gamma-ray spectra were measured with two coaxial Ge(Li) detectors. ``From the individual values the average half-lives were determined to be (9.2$\pm$0.4)~min for $^{160}$Tm and (4.3$\pm$0.2)~min for $^{158}$Tm.'' The quoted half-life for $^{158}$Tm is included in the weighted average for the currently accepted value of 3.98(6)~min.

\subsection*{$^{159}$Tm}\vspace{0.0cm}
In 1971, Ekstr\"om et al.\ identified $^{159}$Tm in the paper entitled ``Nuclear spins of neutron deficient thulium isotopes'' \cite{1971Eks01}. Stable erbium targets were irradiated with 90 MeV protons from the Uppsala synchrocyclotron. $^{159}$Tm was identified with the atomic-beam magnetic resonance method. ``The half-life of T$_{1/2}$=9~min of $^{159}$Tm obtained in our measurements is in reasonable agreement with the value 11$\pm$3~min obtained by Gromov et al. and the 12$\pm$1~min obtained by de Boer et al.'' The quoted half-life is in reasonable agreement with the currently adopted value 9.13(16)~min. The previous measured half-lives of 11(3)~min \cite{1968Gro01} and 12(1)~min \cite{1970deB02} mentioned in the quote were only included in internal reports.

\subsection*{$^{160}$Tm}\vspace{0.0cm}
``New isotopes $^{158}$Tm and $^{160}$Tm'' was published in 1970 documenting the observation of $^{160}$Tm by de Boer et al.\ \cite{1970deB01}. Erbium oxide samples enriched in $^{164}$Er were irradiated with 54 MeV protons from the Amsterdam synchrocyclotron and $^{160}$Tm was populated in (p,5n) reactions. Gamma-ray spectra were measured with two coaxial Ge(Li) detectors. ``From the individual values the average half-lives were determined to be (9.2$\pm$0.4)~min for $^{160}$Tm and (4.3$\pm$0.2)~min for $^{158}$Tm.'' The quoted half-life for $^{160}$Tm is included in the weighted average for the currently accepted value of 9.4(3)~min.

\subsection*{$^{161}$Tm}\vspace{0.0cm}
Harmatz et al.\ reported their observation of $^{161}$Tm in the 1959 paper ``Nuclear spectroscopy of odd-mass (161-173) Nuclides produced by proton irradiation of Er and Yb'' \cite{1959Har01}. Enriched $^{162}$Er targets were irradiated with 12$-$22 MeV proton beams from the Oak Ridge 86-in.\ cyclotron. Conversion electron spectra were measured following chemical separation. ``A target enriched in Er$^{162}$ gave rise to an activity which we assign to Tm$^{161}$. The half-life is 30$\pm$10~minutes and a number of internally converted gamma-ray transitions were observed to follow the electron-capture decay of Tm$^{161}$.'' This half-life agrees with the currently accepted value 30.2(8)~min. A year later a 32-min half-life was assigned to $^{161}$Tm independently by Butement and and Glentworth \cite{1960But02}.

\subsection*{$^{162}$Tm}\vspace{0.0cm}
The observation of $^{162}$Tm was announced by Abdumalikov et al.\ in the 1963 paper ``New Yb$^{162}$ and Tm$^{162}$'' \cite{1963Abd01}. A tantalum target was bombarded with 660 MeV protons and $^{162}$Tm was formed in spallation reactions. Conversion electron spectra were measured with a constant homogeneous magnetic field beta-spectrograph and a triple focusing beta-spectrometer. ``The data obtained are in disagreement with the data of Wilson and Pool: 1) the half-life of Tm$^{172}$ according to our data is 21.5~min and not 77~min; 2) in the decay of Tm$^{162}$ positrons of a sufficient intensity arise which were not noticed by Wilson and Pool.'' This half-life is included in the weighted average to obtain the currently accepted value of 21.70(19)~min. The 77(4)~min half-life by Wilson and Pool \cite{1960Wil03} was evidently incorrect.

\subsection*{$^{163}$Tm}\vspace{0.0cm}
Harmatz et al.\ reported their observation of $^{163}$Tm in the 1959 paper ``Nuclear spectroscopy of odd-mass (161$-$173) Nuclides produced by proton irradiation of Er and Yb'' \cite{1959Har01}. Enriched $^{164}$Er targets were irradiated with 12$-$22 MeV proton beams from the Oak Ridge 86-in.\ cyclotron. Conversion electron spectra were measured following chemical separation. ``Proton irradiation of targets enriched with Er$^{164}$ gave rise to an activity of 2.0-hr half-life which, on the basis of activation data is due to Tm$^{163}$.'' This half-life reasonably agrees with the currently accepted value of 1.810(5)~h. A year later a 2-h half-life was assigned to $^{163}$Tm independently by Butement and Glentworth \cite{1960But02}.

\subsection*{$^{164}$Tm}\vspace{0.0cm}
The 1960 paper ``The decay chain Yb$^{164}\rightarrow$Tm$^{164}\rightarrow$Er$^{164}$'' by Abdurazakov et al.\ reported the discovery of $^{164}$Tm \cite{1960Abd01}. A tantalum target was irradiated with 680 MeV protons from the Dubna synchrocyclotron. $^{164}$Yb was produced in spallation reactions decaying to $^{164}$Tb. Gamma-rays, electrons and positrons were measured following chemical separation. ``The value of the Tm$^{164}$ half-life period, averaged by more than 10 measurements, amounts to 2.0$\pm$0.5~min.'' This half-life is included in the calculation of the currently accepted value of 2.0(1)~min. The decay of $^{164}$Yb to $^{164}$Tm had previously been reported but no details of the decay of $^{164}$Tm were given \cite{1960Abd02}. Later in the year a 2~min half-life was assigned to $^{164}$Tm independently by Dalkhsuren et al.\ \cite{1960Dal01}.

\subsection*{$^{165}$Tm}\vspace{0.0cm}
The discovery of $^{165}$Tm was reported in ``Erbium$^{163}$ and Thulium$^{165}$'' by Handley and Olson in 1953 \cite{1954Han02}. Erbium oxide was bombarded with 24 MeV protons from the ORNL 86-in.\ cyclotron. Decay curves and $\gamma$-ray spectra were measured with a scintillation spectrometer  following chemical separation. ``Therefore, the 24.5-hour activity is assigned to Tm$^{165}$ and it is the parent of the 10.5-hour Er$^{165}$.'' This is within a factor of two of the currently accepted value of 30.06(3)~h. Less than five months later Michel and Templeton independently reported a half-life of 29~hr \cite{1954Mic01}.

\subsection*{$^{166,167}$Tm}\vspace{0.0cm}
Wilkinson and Hicks reported the first observation of $^{166}$Tm and $^{167}$Tm in the 1948 paper ``Some new radioactive isotopes of Tb, Ho, Tm, Lu, Ta, W, and Re'' \cite{1948Wil02}. The Berkeley 60-in.\ cyclotron was used to bombard holmium with 20 and 38~MeV $\alpha$-particles. Absorption measurements were performed and decay curves recorded following chemical separation. The results were only summarized in a table. The measured half-lives of 7.7~h ($^{166}$Tm) and 9~d ($^{167}$Tm) agree with the currently accepted values of 7.70(3)~h and 9.25(2)~d, respectively.

\subsection*{$^{168}$Tm}\vspace{0.0cm}
$^{168}$Tm was identified by Wilkinson and Hicks in the 1949 paper ``Radioactive isotopes of the rare earths. I. Experimental techniques and thulium isotopes'' \cite{1949Wil03}. A holmium target was bombarded with 38 MeV $\alpha$-particles from the 60-in.\ Berkeley cyclotron. Electrons and $\gamma$-rays were measured following chemical separation. ``The latter is reported to have no $\gamma$-radiation, and the x- and $\gamma$-radiation observed in the thulium fraction decays with a half-life of 85 days. The allocation to mass 168 on the basis of reaction yields is thus confirmed.'' The quoted half-life is included in the calculation of the currently adopted value of 93.1(2)~d. Previously a 100~d activity was assigned to either $^{167}$Tm or $^{168}$Tm \cite{1948Wil02}.

\subsection*{$^{169}$Tm}\vspace{0.0cm}
In 1934, Aston reported the first observation of $^{169}$Tm in ``The isotopic constitution and atomic weights of the rare earth elements'' \cite{1934Ast04}. The rare earth elements were analyzed with the Cavendish mass spectrograph. ``Thulium (69) is simple 169.''

\subsection*{$^{170}$Tm}\vspace{0.0cm}
Neuninger and Rona discovered $^{170}$Tm in 1936 as described in ``\"Uber die k\"unstliche Aktivit\"at von Thulium'' \cite{1936Neu01}. Slow neutrons from a radon-beryllium source irradiated a thulium sample and the subsequent decay curve was measured. ``Da wir nun die Aktivit\"at seit einigen Monaten messend verfolgen, k\"onnen wir die Halbwertszeit mit einem Wert von T = 4 Monaten mit einer Genauigkeit von $\pm$1/2 Monaten angeben.'' [Because we now measured the activity for several months, we are able to quote a half-life with a value of T = 4 months and an accuracy of $\pm$1/2 months.] This half-life agrees with the presently adopted value of 128.6(3)~d.

\subsection*{$^{171}$Tm}\vspace{0.0cm}
$^{171}$Tm was first identified by DeBenedetti and McGowan in the 1948 paper ``Short-lived isomers of nuclei'' \cite{1948DeB01}. Radioactive sources of $^{171}$Er were produced by neutron capture in the Oak Ridge pile and delayed coincidences between the $\beta$-particles and $\gamma$-rays were measured with two Geiger counters. ``Out of 60 nuclei investigated, 4 short-lived isomeric states were found. These are: Ta$^{181*}$ (22~$\mu$sec.), Re$^{187*}$ (0.65~$\mu$sec.), Tm$^{169*}$ (1~$\mu$sec.), and Tm$^{161*}$ (2.5~$\mu$sec.).'' The measured half-life for this $^{171}$Tm isomer agrees with the currently adopted value of 2.60(2)~$\mu$s. The first observation of the long-lived ground-state was reported in the same year as a conference abstract \cite{1948Ket01} and only seven years later in the refereed literature \cite{1955Bis01}.

\subsection*{$^{172}$Tm}\vspace{0.0cm}
The discoveries of $^{172}$Tm was announced in 1956 by Nethaway et al.\ in the paper ``New isotopes: Er$^{172}$ and Tm$^{172}$'' \cite{1956Net01}. Erbium oxide was irradiated with neutrons from the Idaho Materials Testing Reactor, and $^{172}$Tm was populated by $\beta$-decay from $^{152}$Er which was formed by sequential neutron capture from $^{170}$Er. Decay curves were measured following chemical separation. ``Two new isotopes, Er$^{172}$ of half-life 49.8$\pm$1~hr, and Tm$^{172}$ of half-life 63.6$\pm$0.3~hr, have been found.'' This half-life is used in the calculation of the currently accepted value of 63.6(2)~h.

\subsection*{$^{173}$Tm}\vspace{0.0cm}
Kuroyanagi et al.\ observed $^{173}$Tm in 1961 as reported in ``New activities in rare earth region produced by the ($\gamma$,p) reactions'' \cite{1961Kur01}. Ytterbium oxide powder was irradiated with $\gamma$-rays at the Tohoku 25 MeV betatron. Decay curves were measured with a beta ray analyzer or an end-window G-M counter and $\beta$-ray spectra were recorded with a plastic scintillator. ``The 7.2-h activity is assigned to Tm$^{173}$ from its decay properties and the ($\gamma$,p) yield ratios to such well known activity as Tm$^{172}$.'' This half-life is close to the currently accepted value of 8.24(8)~h.

\subsection*{$^{174}$Tm}\vspace{0.0cm}
$^{174}$Tm was first observed by Wille and Fink in 1960 as reported in ``Activation cross sections for 14.8-Mev neutrons and some new radioactive nuclides in the rare earth region'' \cite{1960Wil02}. An ytterbium target enriched in $^{174}$Yb was irradiated with neutrons produced in the $^{3}$H(d,n)$^{4}$He reaction from the University of Arkansas Cockcroft-Walton accelerator. Decay curves were measured with an aluminum-walled, methane-flow beta-proportional counter and $\gamma$-spectra were measured with a Na(Tl) detector. ``From 98.4\% enriched Yb$^{174}$, the 2.0-min activity was not found, but a new 5.5$\pm$0.5~min half-life was observed, probably due to Tm$^{174}$ from the (n,p) reaction.''  This half-life was included in the calculation of the currently accepted value of 5.4(1)~min.

\subsection*{$^{175}$Tm}\vspace{0.0cm}
Kuroyanagi et al.\ observed $^{175}$Tm in 1961 as reported in ``New activities in rare earth region produced by the ($\gamma$,p) reactions'' \cite{1961Kur01}. Pure oxide powder was irradiated with $\gamma$-rays at the Tohoku 25 MeV betatron. Decay curves were measured with a beta ray analyser or an end-window G-M counter and $\beta$-ray spectra were recorded with a plastic scintillator. ``From the decay characteristics and the yield of the 20-min activity, it is considered to be assigned to Tm$^{175}$.'' This half-life is close to the currently accepted value of 15.2(5)~min. A previously reported half-life of 19~min was assigned to either $^{172}$Tm, $^{173}$Tm, or $^{175}$Tm \cite{1950But01}

\subsection*{$^{176}$Tm}\vspace{0.0cm}
The observation of $^{176}$Tm was described by Takashi et al.\ in the 1961 paper ``Some new activities produced by fast neutron bombardments \cite{1961Tak01}. Fast neutrons produced by bombarding graphite targets with 20 MeV deuterons from the Tokyo 160 cm variable energy cyclotron irradiated a ytterbium oxide sample. Gamma- and beta-ray spectra were measured with NaI(Tl) and plastic scintillators, respectively. ``A previously unknown activity of 1.5~min was observed. Similar value of half-life was found in a beta spectrum measurement. Accordingly one may tentatively assign this new isotope to Tm$^{176}$.'' The reported half-life agrees with the presently adopted value of 1.9(1)~min.

\subsection*{$^{177}$Tm}\vspace{0.0cm}
$^{177}$Tm was identified by Rykaczewski et al.\ as reported in ``Investigation of neutron-rich rare-earth nuclei including the new isotopes  $^{177}$Tm and $^{184}$Lu'' in 1989 \cite{1989Ryk01}. A stack of tungsten and tantalum foils were bombarded with 9$-$15~MeV/u $^{136}$Xe, $^{186}$W, and $^{238}$U beams from the GSI UNILAC accelerator. Plastic scintillators and Ge(Li) detectors were used to measure $\beta$- and $\gamma$-ray spectra, respectively following on-line mass separation. ``Taking into account the results from all these experiments the half-life value of $^{177}$Tm is determined to be 85$^{+10} _{-15}$s.'' This half-life agrees with the presently adopted value of 90(6)~s.


\section{Discovery of $^{149-180}$Yb}
Thirty-one ytterbium isotopes from A = 149$-$180 have been discovered so far; these include 7 stable, 19 neutron-deficient and 5 neutron-rich isotopes. According to the HFB-14 model \cite{2007Gor01}, $^{231}$Yb should be the last odd-even particle stable neutron-rich nucleus while the even-even particle stable neutron-rich nuclei should continue at least through $^{234}$Yb. On the neutron deficient side five more isotopes should be particle bound ($^{145-148,150}$Yb). In addition, seven more isotopes ($^{138-144}$Yb) could still have half-lives longer than 10$^{-9}$~s \cite{2004Tho01}. Thus, about 66 isotopes have yet to be discovered corresponding to 68\% of all possible ytterbium isotopes.

\begin{figure}
	\centering
	\includegraphics[scale=.7]{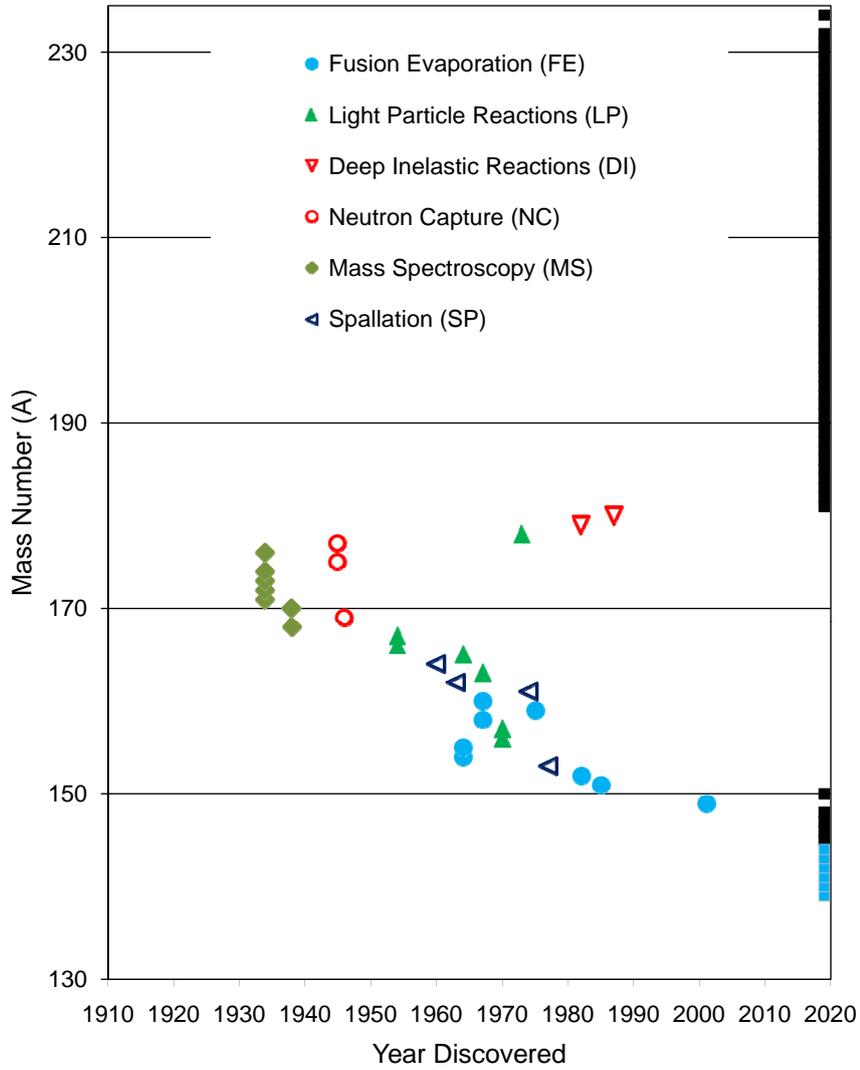}
	\caption{Ytterbium isotopes as a function of time when they were discovered. The different production methods are indicated. The solid black squares on the right hand side of the plot are isotopes predicted to be bound by the HFB-14 model. On the proton-rich side the light blue squares correspond to unbound isotopes predicted to have half-lives larger than $\sim 10^{-9}$~s.}
\label{f:year-yb}
\end{figure}

Figure \ref{f:year-yb} summarizes the year of discovery for all ytterbium isotopes identified by the method of discovery. The range of isotopes predicted to exist is indicated on the right side of the figure. The radioactive ytterbium isotopes were produced using fusion evaporation reactions (FE), light-particle reactions (LP), deep-inelastic reactions (DI), neutron capture (NC), and spallation (SP). The stable isotopes were identified using mass spectroscopy (MS). The discovery of each ytterbium isotope is discussed in detail and a summary is presented in Table 1.

\subsection*{$^{149}$Yb}\vspace{0.0cm}
The discovery of $^{149}$Yb by Xu et al.\ was announced in the 2001 paper ``New $\beta$-delayed proton precursor $^{149}$Yb near the proton drip line'' \cite{2001Xu01}. A 232 MeV $^{40}$Ca beam from the Lanzhou sector-focusing cyclotron bombarded an enriched $^{122}$Sn target and $^{149}$Yb was formed in the $^{122}$Sn($^{40}$Ca,3n) fusion-evaporation reaction. A helium-jet fast tape transport system was used to move the recoils in front of silicon surface barrier and HPGe detectors for protons-$\gamma$-ray coincidence measurements. ``The decay curve of the 647~keV $\gamma$ line coincident with 2.5-6.4~MeV protons is shown in the inset of [the figure] from which the half-life of $^{149}$Yb was extracted to be 0.7$\pm$0.2~s.'' The quoted half-life is the currently accepted value.

\subsection*{$^{151}$Yb}\vspace{0.0cm}
In 1985 Kleinheinz et al.\ reported the first observation of $^{151}$Yb in ``Beta-decay of $^{151}$Yb'' \cite{1985Kle01}. A $^{86}$Ru target was bombarded with a $^{58}$Ni beam and $^{151}$Yb was populated in the (2pn) fusion evaporation reaction. Recoil products were separated on-line and $^{151}$Yb was identified by measuring $\gamma$- and X-rays. ``The most precise result is obtained from the X-ray data which give T$_{1/2}$($^{151}$Yb)=1.6(2)~s.'' This half-life agrees with the currently accepted value 1.6(1)~s.

\subsection*{$^{152}$Yb}\vspace{0.0cm}
$^{152}$Yb was observed by Nolte et al.\ and results were published in the paper ``Seniority isomerism in the N=82 isotone $^{152}$Yb; favoured $\beta$ transitions $\pi h_{11/2} \rightarrow \nu h_{9/2}$'' \cite{1982Nol01}. An enriched $^{96}$Ru target was irradiated with 238 and 250~MeV $^{58}$Ni from the Munich MP tandem and postaccelerator and $^{152}$Yb was formed in the $^{96}$Ru($^{58}$Ni,2p) fusion-evaporation reaction. Gamma-ray singles and $\gamma-\gamma$ coincidences were measured with  a coaxial Ge(Li) detector and a planar Ge detector. ``$\beta$-decay schemes of $^{152}$Yb (3.2$\pm$0.3~s), $^{152}$Tm$^{(m)}$(8.0$\pm$1.0~s) and $^{151}$Tm (3.8$\pm$0.8~s) have been derived.'' This half-life for $^{152}$Yb is included in the calculation of the currently accepted value of 3.04(6)~s. In an earlier report the existence of $^{152}$Yb was inferred from $\alpha$-correlation measurements: ``Further, correlations were measured between the $\alpha$ lines of $^{157}$Ta$ - ^{153}$Tm and $^{156}$Hf$ - ^{152}$Er that prove a $\beta$-decay of the new isotopes $^{153}$Lu, $^{152}$Yb, and $^{152}$Tm.'' \cite{1981Hof01}. However, no properties of $^{152}$Yb or its decay were measured.

\subsection*{$^{153}$Yb}\vspace{0.0cm}
The observation of $^{153}$Yb was reported by Hagberg et al.\ in the 1977 paper ``Alpha decay of neutron-deficient ytterbium isotopes and their daughters'' \cite{1977Hag02}. The CERN synchro-cyclotron was used to bombard tantalum with 600~MeV protons. $^{153}$Yb was separated with the ISOLDE on-line mass separator facility and $\alpha$ particles were measured with two silicon surface-barrier detectors. $^{153}$Yb was identified by the growth preceding the decay of $\alpha$ particles from $^{153}$Tm. ``The decay data from this experiment, shown in [the figure], determined the half-life of $^{153}$Yb to be 4.0$\pm$0.5~s.'' The quoted half-life is included in the calculation of the currently accepted value of 4.2(2)~s.

\subsection*{$^{154,155}$Yb}\vspace{0.0cm}
``Alpha-decay properties of some thulium and ytterbium isotopes near the 82-neutron shell'' by Macfarlane announced the discovery of $^{154}$Yb and $^{155}$Yb in 1964 \cite{1964Mac01}. Samarium oxide (enriched in $^{144}$Sm) and neodymium oxide (enriched in $^{142}$Nd) were bombarded with 106$-$151 MeV $^{16}$O and 131$-$195 MeV $^{20}$Ne beams from the Berkeley heavy-ion linear accelerator Hilac, respectively. $^{154}$Yb and $^{155}$Yb were formed in (xn) fusion evaporation reactions and identified by measuring excitation functions and $\alpha$-decay spectra. ``One of the Yb alpha emitters has an alpha-particle energy of 5.33~MeV and decays with a half-life of 0.39~sec... This result strongly suggests that this new nuclide is probably Yb$^{154}$... The second Yb alpha activity that was observed has an alpha-particle energy of 5.21~MeV, and decays with a half-life of 1.65~sec. The assignment of this activity to Yb$^{155}$ was made by a procedure used for Yb$^{154}$, making use of excitation function data from previous work to identify the reaction producing the activity.'' These half-lives of 0.39(4)~s and 1.65(15)~s agree with the currently adopted values of 0.409(2)~s and 1.793(13)~s for $^{154}$Yb and $^{155}$Yb, respectively.

\subsection*{$^{156,157}$Yb}\vspace{0.0cm}
The discovery of $^{156}$Yb and $^{157}$Yb was reported in ``Production of rare-earth $\alpha$ emitters with energetic $^{3}$He particles; new isotopes: $^{151}$Er, $^{156}$Yb, and $^{157}$Yb'' by Toth et al.\ in 1970 \cite{1970Tot01}. Erbium oxide targets enriched in $^{162}$Er were bombarded with a 102.1~MeV $^3$He beam from the Oak Ridge isochronous cyclotron (ORIC). $^{156}$Yb and $^{157}$Yb were produced in (9n) and (8n) reactions, respectively. Recoils were transported to a Si(Au) detector with a helium gas transport system where $\alpha$-decay spectra were measured. ``Least-squares analysis indicated a genetic relationship between two radioactive components, one with a half-life of 24$\pm$1~sec and the other with a 9.8-sec half-life. Because this latter value is that of $^{152}$Er, the parent-daughter relationship establishes the existence of a new ytterbium nuclide, $^{156}$Yb... The only calculated curve that was consistent with the 4.50-MeV data points was the one for a ($^3$He,8n) reaction. There is a strong indication then that the 4.50-MeV $\alpha$ group is due to the decay of $^{157}$Yb.'' The reported half-lives of 24(1)~s and 34(3)~s agree with the currently accepted values of  26.1(7)~s and 38.6(10)~s, for $^{156}$Yb and $^{157}$Yb, respectively.

\subsection*{$^{158}$Yb}\vspace{0.0cm}
Ward et al.\ published the identification of $^{158}$Yb in the 1967 paper ``Gamma rays following $^{40}$Ar-induced reactions'' \cite{1967War02}. Isotopically enriched tellurium targets were bombarded with $^{40}$Ar beams and ytterbium isotopes were populated in (xn) fusion-evaporation reactions. Gamma-ray spectra were studied using a lithium-drifted germanium counter. ``We have bombarded separated Sn and Te isotopes with $^{40}$Ar projectiles in order to study the ($^{40}$Ar,xn) reactions and evaluated them as a means to produce excited nuclei for spectroscopic studies. This proves to be an excellent method for populating ground-band collective levels, and such levels have been identified as the 88-, 90-, and 92-neutron Er and Yb isotopes.'' The first three transition of the rotational ground-state band were measured for $^{158}$Yb.

\subsection*{$^{159}$Yb}\vspace{0.0cm}
$^{159}$Yb was observed in 1975 by Trautmann et al.\ and published in the paper ``Spectroscopy on the gamma decay of highly excited high-spin states by angular-correlation and feeding-time measurements'' \cite{1975Tra01}. $^{18}$O accelerated to 74.7 MeV by the Munich MP tandem Van de Graaff irradiated a $^{144}$Sm target and $^{159}$Yb was populated in the (3n) fusion evaporation reaction. Gamma-ray spectra were measured with a NaI(Tl) $\gamma$ spectrometer and a Ge(Li) detector. ``Unfortunately, our values for t$_{1/2}^{feed}$ in $^{159}$Yb are only upper limits, also the intensity of E2 transitions has to be measured yet for this nucleus.'' Five transitions of the $i_{13/2}$ band in $^{159}$Yb were measured. The first measurement of the half-life of $^{159}$Yb was mentioned around the same time \cite{1976Gro01}, however, it was based on a conference report \cite{1975Gon01}.

\subsection*{$^{160}$Yb}\vspace{0.0cm}
Ward et al.\ published the identification of $^{160}$Yb in the 1967 paper ``Gamma rays following $^{40}$Ar-induced reactions'' \cite{1967War02}. Isotopically enriched tellurium targets were bombarded with $^{40}$Ar beams and ytterbium isotopes were populated in (xn) fusion-evaporation reactions. Gamma-ray spectra were studied using a lithium-drifted germanium counter. ``We have bombarded separated Sn and Te isotopes with $^{40}$Ar projectiles in order to study the ($^{40}$Ar,xn) reactions and evaluated them as a means to produce excited nuclei for spectroscopic studies. This proves to be an excellent method for populating ground-band collective levels, and such levels have been identified as the 88-, 90-, and 92-neutron Er and Yb isotopes.'' The first five transition of the rotational ground-state band were measured for $^{160}$Yb.

\subsection*{$^{161}$Yb}\vspace{0.0cm}
``Method for obtaining separated short-lived isotopes of rare earth elements'' was published in 1974 by Latuszynski et al. documenting their observation of $^{161}$Yb \cite{1974Lat02}. A tantalum target was bombarded with 660~MeV protons from the Dubna synchrocyclotron. Gamma-ray spectra and decay curves were measured at the end of an electromagnetic separator. ``Using the method proposed for investigations in the field of nuclear spectroscopy the gamma-spectra of short-living isotopes with T$_{1/2} \le$ 1 minute have been measured. The new isotopes $^{161}$Yb (4.2~min), $^{148}$Dy (3.5~min) $^{132}$Pr (1.6~min) have been identified.'' The observed half-life corresponds to the present value. The previous assignment of a 82(4)~min half-life to $^{161}$Yb \cite{1959Kal01} was evidently incorrect.

\subsection*{$^{162}$Yb}\vspace{0.0cm}
The observation of $^{162}$Yb was announced by Abdumalikov et al.\ in the 1963 paper ``New Yb$^{162}$ and Tm$^{162}$'' \cite{1963Abd01}. A tantalum target was bombarded with 660 MeV protons and $^{162}$Yb was formed in spallation reactions. Conversion electron spectra were measured with a constant homogeneous magnetic field beta-spectrograph and a triple focusing beta-spectrometer. ``The half life of Yb$^{162}$ appears to be somewhat smaller than 26~min but larger than 21.5~min.'' This half-life range is close to the presently adopted value of 18.87(19)~min.

\subsection*{$^{163}$Yb}\vspace{0.0cm}
The first observation of $^{163}$Yb was reported by Paris in ``La p\'eriode de d\'ecroissance de l'ytterbium 163'' in 1967 \cite{1967Par01}. A Tm$_2$O$_3$ target was bombarded with protons from the Orsay synchrocyclotron and $^{163}$Yb was populated in the (p,7n) reaction. Gamma-ray spectra were measured with a germanium detector following element and mass separation. ``La d\'esint\'egration $^{163}$Yb$\rightarrow^{163}$Tm a \'et\'e observ\'ee pour la premi\`ere fois. La d\'etermination de la d\'ecroissance de plusieurs sources s\'epar\'ees isotopiquement permet d'attribuer \`a $^{163}$Yb une p\'eriode T$_{1/2}$ = 10,9$\pm$0,5~mn.'' [The decay $^{163}$Yb$\rightarrow^{163}$Tm was observed for the first time. The determination of the decay of several isotopically separated sources was used to assign a half-life of T$_{1/2}$ = 10.9$\pm$0.5~min to $^{163}$Yb.] This half-life agrees with the presently adopted value of 11.05(25)~min.

\subsection*{$^{164}$Yb}\vspace{0.0cm}
In ``New radioactive isotopes of the rare earth elements'' Butement and Glentworth reported the discovery of $^{164}$Yb in 1959 \cite{1960But02}. A Tm$_2$O$_3$ target was bombarded by 230~MeV protons and $^{164}$Yb was produced in spallation reactions. Decay curves were measured with a Geiger counter and $\gamma$-ray spectra were recorded with a scintillation spectrometer following chemical separation. ``The most probable mass assignment of the 85~min activity is to $^{164}$Yb.'' This half-life is close to the currently accepted value of 75.8(17)~min. Less than four months later Abdurazakov et al.\ independently reported a half-life of 75(2)~min \cite{1960Abd01,1960Abd02}. Previously, half-lives of 82(4)~min and 74~min were assigned to $^{161}$Yb \cite{1959Kal01} and $^{167}$Yb \cite{1955Ner01}, respectively.

\subsection*{$^{165}$Yb}\vspace{0.0cm}
$^{165}$Yb was first observed by Paris as described in the 1964 paper ``D\'etermination des p\'eriodes des ytterbiums 165 et 164'' \cite{1964Par01}. Thulium oxide was irradiated with protons from the Orsay synchrocyclotron. $^{165}$Yb was populated in (p,5n) reactions and identified by measuring decay curves after the first stage of an isotope separator. ``La figure montre la d\'ecroissance de $^{163}$Yb, observ\'ee a l'aide d'un compteur $\gamma$ \`a scintillations suivi d'un discrimanteur et d'une \'echelle... La p\'eriode obtenue est T$_{1/2}$ = 10,5$\pm$0,5~mn, beaucoup plus courte que celle admise habituellement (74 ou 82 mn).''  [The figure shows the decay of $^{163}$Yb, observed using a $\gamma$ scintillation counter followed by a discriminator and a scaler... The obtained period T$_{1/2}$ =10.5$\pm$0.5~min is much shorter than usually accepted (74 or 82 min).] This half-life agrees with the presently adopted value of 9.9(3)~min. The 74 or 82~min half-lives quoted refer to previous measurements which had originally been assigned to $^{161}$Yb \cite{1959Kal01} and $^{167}$Yb \cite{1955Ner01}, respectively.

\subsection*{$^{166}$Yb}\vspace{0.0cm}
``Mass assignments by isotope separation'' was published isn 1954 by Michel and Templeton documenting the observation of $^{166}$Yb \cite{1954Mic01}. The Berkeley 184-inch cyclotron was used to produce radioactive isotopes which were identified with a time-of-flight isotope separator. The resulting activities were measured with a G-M counter and a scintillation spectrograph. ``However, the following previously known isotopes have been assigned in the course of this work (half-lives are by direct observation of the decay of separated samples): Yb$^{166}$ (58$\pm$1~hours), Yb$^{169}$ (32~days), Tm$^{166}$ (7.7~hours), Tm$^{156}$ (9.6)~days.'' The quoted half-life of$^{166}$Yb agrees with the currently accepted value of 56.7(1)~h.

\subsection*{$^{167}$Yb}\vspace{0.0cm}
In ``Ytterbium-167'' Handley and Olson reported their 1954 discovery of $^{167}$Yb \cite{1954Han01}. Tm$_{2}$O$_{3}$ was irradiated with 24 MeV protons from the ORNL 86-in.\ cyclotron. Decay curves were measured with a GM tube and $\gamma$-ray spectra were recorded with a NaI(Tl) scintillation spectrometer following chemical separation. ``Analysis of the Yb peak by decay curves and gamma spectra reveals three components: 18.5-minute Yb$^{167}$, 33-day Yb$^{169}$, and 9.4-day Tm$^{167}$.'' This half-life for $^{167}$Yb is close to the currently accepted value, 17.5(2)~min.

\subsection*{$^{168}$Yb}\vspace{0.0cm}
Dempster reported the discovery of $^{168}$Yb in the 1938 paper ``The isotopic constitution of gadolinium, dysprosium, erbium and ytterbium'' \cite{1938Dem01}. An ytterbium oxide sample reduced with neodymium was used for analysis in the Chicago mass spectrograph. ``The ytterbium oxide was reduced by neodymium and also showed two new isotopes at masses 170 and 168.''

\subsection*{$^{169}$Yb}\vspace{0.0cm}
The observation of $^{169}$Yb was reported by Bothe in 1946 in ``Die Aktivierung der seltenen Erden durch thermische Neutronen I'' \cite{1946Bot01}. Ytterbium oxide was irradiated with thermal neutrons produced by the bombardment of beryllium with deuterons. Decay and absorption curves were measured. ``Hiernach beruht die 33-d-Aktivit\"at auf K-Einfang; sie ist nicht mit $\beta$-Strahlung verbunden, wie oben gezeigt. Damit ist die Zuordnung zu Yb$^{169}$ eindeutig.'' [The 33-d activity is therefore due to K-capture; there is no related $\beta$-radiation as shown above. Thus, the assignment to Yb$^{169}$ is clear.] This half-life agrees with the present value of 32.026(5)~d.

\subsection*{$^{170}$Yb}\vspace{0.0cm}
Dempster reported the discovery of $^{170}$Yb in the 1938 paper ``The isotopic constitution of gadolinium, dysprosium, erbium and ytterbium'' \cite{1938Dem01}. An ytterbium oxide sample reduced with neodymium was used for analysis in the Chicago mass spectrograph. ``The ytterbium oxide was reduced by neodymium and also showed two new isotopes at masses 170 and 168.''

\subsection*{$^{171-174}$Yb}\vspace{0.0cm}
In 1934, Aston reported the first observation of $^{171}$Yb, $^{172}$Yb, $^{173}$Yb, and $^{174}$Yb in ``The isotopic constitution and atomic weights of the rare earth elements'' \cite{1934Ast04}. Rare earth elements were analyzed with the Cavendish mass spectrograph. ``Ytterbium (70) appears to contain mass numbers 171,172,173, 174, 176, of which 174 is the strongest.''

\subsection*{$^{175}$Yb}\vspace{0.0cm}
The identification of $^{175}$Yb was reported by Atterling et al.\ in 1945 in ``Neutron-induced radioactivity in lutecium and ytterbium'' \cite{1945Att01}. Ytterbium samples were bombarded with fast and slow neutrons produced by bombarding LiOH with 6~MeV deuterons and beryllium with 6.5 MeV deuterons from the Stockholm cyclotron, respectively. The resulting activities were measured with a Wulf string electrometer and a Geiger-M\"uller counter. ``An initial slope of 4.2 would mean that the mother substance had a half-life of only 2.6~d and the decay curve would soon approach the 6.6~d period. We therefore assign the 4.2~d period to Yb$^{175}$.'' The reported half-life agrees with the currently accepted value 4.185(1)~d.

\subsection*{$^{176}$Yb}\vspace{0.0cm}
In 1934, Aston reported the first observation of $^{176}$Yb in ``The isotopic constitution and atomic weights of the rare earth elements'' \cite{1934Ast04}. \cite{1934Ast04}. Rare earth elements were analyzed with the Cavendish mass spectrograph. ``Ytterbium (70) appears to contain mass numbers 171,172,173, 174, 176, of which 174 is the strongest.''

\subsection*{$^{177}$Yb}\vspace{0.0cm}
The identification of $^{177}$Yb was reported by Atterling et al.\ in 1945 in ``Neutron-induced radioactivity in lutecium and ytterbium'' \cite{1945Att01}. Ytterbium samples were bombarded with fast and slow neutrons produced by bombarding LiOH with 6~MeV deuterons and beryllium with 6.5 MeV deuterons from the Stockholm cyclotron, respectively. The resulting activities were measured with a Wulf string electrometer and a Geiger-M\"uller counter. ``As the cross-section for the 1.9~h period is very small we can hardly expect to find the 6.6~d period in the decay curve of Yb. We therefore assign the 4.2~d period to Yb$^{175}$ and the 1.9~h period to Yb$^{177}$.'' The reported half-life agrees with the currently accepted value 1.911(3)~h.

\subsection*{$^{178}$Yb}\vspace{0.0cm}
$^{178}$Yb was identified in 1973 by Orth et al.\ and published in ``Decay of $^{178}$Yb and the isomers of $^{178}$Lu'' \cite{1973Ort01}. Enriched $^{176}$Yb targets were bombarded with 12-MeV tritons from the Los Alamos Van de Graaff accelerator and $^{178}$Yb was formed in the ($^3$H,n) reaction. Gamma-and beta-ray spectra were measured following chemical separation. ``Our value for the half-life of $^{178}$Yb is, therefore, 74$\pm$3~min.'' The quoted half-life is the currently adopted value.

\subsection*{$^{179}$Yb}\vspace{0.0cm}
$^{179}$Yb was discovered by Kirchner et al.\ in 1981 and reported in ``New neutron-rich $^{179}$Yb and $^{181,182}$Lu isotopes produced in reactions of 9~MeV/u $^{136}$Xe ions on tantalum and tungsten targets'' \cite{1982Kir01}. A $^{136}$Xe beam from the GSI UNILAC accelerator bombarded tungsten and tantalum targets. $^{179}$Yb was identified with an online-mass separator and $\beta$-, $\gamma$-, and X-ray decay spectroscopy. ``[The figure] shows a part of the $\gamma$-ray spectrum measured for mass A = 179. The lines assigned to the $^{179}$Yb decay are labelled with energies. The time-analyses of $\beta$-rays (after correction for lutetium $\beta$-rays) and the 612.5~keV $\gamma$-transition agreed within a half-life value of T$_{1/2}$=8.1$\pm$0.8~min.'' This half-life is included in the calculation of the currently accepted value, 8.0(4)~min.

\subsection*{$^{180}$Yb}\vspace{0.0cm}
In 1987, Runte et al.\ identified $^{180}$Yb in the paper entitled ``The decay of the new isotope $^{180}$Yb and the search for the r-process path to $^{180m}$Ta'' \cite{1987Run01}. A 15 MeV/u $^{186}$W beam from the GSI UNILAC bombarded a tungsten/tantalum target and $^{180}$Yb was produced in multi-nucleon transfer reactions. Beta-, gamma- and X-ray spectra of recoil products mass separated with the GSI on-line mass separator were measured. ``Irradiating $^{nat}$W Ta targets with $^{186}$W, the decay of the new isotope $^{180}$Yb was observed and its half-life determined to be 2.4(5)~min.'' This half-life is the currently accepted half-life.

\section{Summary}
The discoveries of the known dysprosium, holmium, erbium, thulium and ytterbium isotopes have been compiled and the methods of their production discussed.
The identification of the isotopes of these elements was relatively easy with the exception of holmium where five isotopes ($^{161-163}$Ho, $^{166}$Ho, and $^{169}$Ho) were initially identified incorrectly. In the other elements, only the half-lives of $^{154,155}$Dy, $^{169}$Er, $^{173}$Er, $^{162}$Tm and $^{161}$Yb were at first not correct. In addition, the half-lives of several isotopes were initially reported with no or only uncertain mass assignments. These isotopes were: $^{151-152}$Dy, $^{159}$Dy, $^{171}$Ho, $^{150}$Er, $^{165}$Er, $^{168}$Tm, $^{175}$Tm, and $^{164}$Yb.

\ack

This work was supported by the National Science Foundation under grants No. PHY06-06007 (NSCL) and PHY10-62410 (REU).

\bibliography{../isotope-discovery-references}

\newpage

\newpage

\TableExplanation

\bigskip
\renewcommand{\arraystretch}{1.0}

\section{Table 1.\label{tbl1te} Discovery of dysprosium, holmium, erbium, thulium and ytterbium isotopes }
\begin{tabular*}{0.95\textwidth}{@{}@{\extracolsep{\fill}}lp{5.5in}@{}}
\multicolumn{2}{p{0.95\textwidth}}{ }\\

Isotope & Dysprosium, holmium, erbium, thulium and ytterbium isotope \\
Author & First author of refereed publication \\
Journal & Journal of publication \\
Ref. & Reference \\
Method & Production method used in the discovery: \\

  & FE: fusion evaporation \\
  & LP: light-particle reactions (including neutrons) \\
  & MS: mass spectroscopy \\
  & NC: neutron capture reactions \\
  & PN: photo-nuclear reactions \\
  & DI: deep-inelastic reactions \\
  & SF: spontaneous fission \\
  & SP: spallation \\

Laboratory & Laboratory where the experiment was performed\\
Country & Country of laboratory\\
Year & Year of discovery \\
\end{tabular*}
\label{tableI}

\datatables 



\setlength{\LTleft}{0pt}
\setlength{\LTright}{0pt}


\setlength{\tabcolsep}{0.5\tabcolsep}

\renewcommand{\arraystretch}{1.0}

\footnotesize 

\begin{longtable}{@{\extracolsep\fill}llllllll@{}}
\caption{Discovery of dysprosium, holmium, erbium, thulium and ytterbium isotopes. See page\ \pageref{tbl1te} for Explanation of Tables}
Isotope & Author & Journal & Ref. & Method & Laboratory & Country & Year\\
\hline\\
\endfirsthead\\
\caption[]{(continued)}
Isotope & Author & Journal & Ref. & Method & Laboratory & Country & Year\\
\hline\\
\endhead
$^{139}$Dy& S.-W. Xu & Phys. Rev. C &\cite{1999Xu01}& FE & Lanzhou & China &1999 \\
$^{140}$Dy & W. Krolas & Phys. Rev. C &\cite{2002Kro01}& FE & Oak Ridge & USA &2002 \\
$^{141}$Dy & J.M. Nitschke & Nucl. Phys. A &\cite{1984Nit01}& FE & Berkeley & USA &1984 \\
$^{142}$Dy & P.A. Wilmarth & Z. Phys. A &\cite{1986Wil01}& FE & Berkeley & USA &1986 \\
$^{143}$Dy & J.M. Nitschke & Z. Phys. A &\cite{1983Nit01}& FE & Berkeley & USA &1983 \\
$^{144}$Dy & N. Redon & Z. Phys. A &\cite{1986Red01}& FE & Grenoble & France &1986 \\
$^{145}$Dy & E. Nolte & Z. Phys. A &\cite{1982Nol02}& FE & Darmstadt & Germany &1982 \\
           & G.D.Alkhazov & Z. Phys. A &\cite{1982Alk01}& SP & Leningrad & Russia &1982 \\
$^{146}$Dy & G.D.Alkhazov & Acta Phys. Pol. B &\cite{1981Alk03}& SP & Leningrad & Russia &1981 \\
$^{147}$Dy & K.S. Toth & Phys. Lett. B &\cite{1975Tot02}& FE & Oak Ridge & USA &1975 \\
$^{148}$Dy & A. Latuszynski & Nukleonika &\cite{1974Lat02}& SP & Dubna & Russia &1974 \\
$^{149}$Dy & K.S. Toth & Phys. Rev. &\cite{1958Tot02}& FE & Berkeley & USA &1958 \\
$^{150}$Dy & K.S. Toth & J. Inorg. Nucl. Chem. &\cite{1959Tot03}& FE & Berkeley & USA &1959 \\
$^{151}$Dy & K.S. Toth & J. Inorg. Nucl. Chem. &\cite{1959Tot03}& FE & Berkeley & USA &1959 \\
$^{152}$Dy & K.S. Toth & Phys. Rev. &\cite{1958Tot02}& FE & Berkeley & USA &1958 \\
$^{153}$Dy & K.S. Toth & Phys. Rev. &\cite{1958Tot02}& FE & Berkeley & USA &1958 \\
$^{154}$Dy & R.D. Macfarlane & J. Inorg. Nucl. Chem. &\cite{1961Mac01}& LP & Berkeley & USA &1961 \\
$^{155}$Dy & K.S. Toth & Phys. Rev. &\cite{1958Tot02}& FE & Berkeley & USA &1958 \\
$^{156}$Dy & D.C. Hess& Phys. Rev. &\cite{1948Hes01}& MS & Argonne & USA &1948 \\
$^{157}$Dy & T.H. Handley & Phys. Rev. &\cite{1953Han02}& LP & Oak Ridge & USA &1953 \\
$^{158}$Dy & A.J. Dempster & Phys. Rev. &\cite{1938Dem01}& MS & Chicago & USA &1938 \\
$^{159}$Dy & F.D.S. Butement & Proc. Phys. Soc. A &\cite{1951But02}& NC & Harwell & UK &1951 \\
$^{160}$Dy & A.J. Dempster & Phys. Rev. &\cite{1938Dem01}& MS & Chicago & USA &1938 \\
$^{161}$Dy & F.W. Aston & Nature &\cite{1934Ast04}& MS & Cambridge & UK &1934 \\
$^{162}$Dy & F.W. Aston & Nature &\cite{1934Ast04}& MS & Cambridge & UK &1934 \\
$^{163}$Dy & F.W. Aston & Nature &\cite{1934Ast04}& MS & Cambridge & UK &1934 \\
$^{164}$Dy & F.W. Aston & Nature &\cite{1934Ast04}& MS & Cambridge & UK &1934 \\
$^{165}$Dy & J. K. Marsh & Nature &\cite{1935Mar01}& NC & Oxford & UK &1935 \\
           & G. Hevesy & Nature &\cite{1935Hev01}& NC & Copenhagen & Denmark &1935 \\
$^{166}$Dy & B.H. Ketelle & Phys. Rev. &\cite{1949Ket01}& NC & Oak Ridge & USA &1949 \\
$^{167}$Dy & R.G. Wille & Phys. Rev. &\cite{1960Wil02}& LP & Arkansas & USA &1960 \\
$^{168}$Dy & R.J. Gehrke& Z. Phys. A &\cite{1982Geh01}& SF & Idaho Falls & USA &1982 \\
$^{169}$Dy & R.M. Chasteler & Phys. Rev. C &\cite{1990Cha01}& DI & Berkeley & USA &1990 \\
 & & & & & & & \\
 & & & & & & & \\
$^{140}$Ho& K. Rykaczewski & Phys. Rev. C &\cite{1999Ryk01}& FE & Oak Ridge & USA &1999 \\
$^{141}$Ho & C.N. Davids & Phys. Rev. Lett. &\cite{1998Dav01}& FE & Argonne & USA &1998 \\
$^{142}$Ho & S.-W. Xu & Phys. Rev. C &\cite{2001Xu02}& FE & Lanzhou & China &2001 \\
$^{143}$Ho & & & & & & & \\
$^{144}$Ho & P.A. Wilmarth & Z. Phys. A &\cite{1986Wil01}& FE & Berkeley & USA &1986 \\
$^{145}$Ho & L. Goettig & Nucl. Phys. A &\cite{1987Goe01}& FE & Daresbury & UK &1987 \\
$^{146}$Ho & S.Z. Gui & Z. Phys. A &\cite{1982Gui01}& FE & Munich & Germany &1982 \\
$^{147}$Ho & E. Nolte & Z. Phys. A &\cite{1982Nol02}& FE & Munich & Germany &1982 \\
$^{148}$Ho & K.S. Toth & Phys. Rev. C &\cite{1979Tot01}& FE & Texas A\&M & USA &1979 \\
$^{149}$Ho & K.S. Toth & Phys. Rev. C &\cite{1979Tot01}& FE & Texas A\&M & USA &1979 \\
$^{150}$Ho & R.D. Macfarlane & Phys. Rev. &\cite{1963Mac02}& FE & Berkeley & USA &1963 \\
$^{151}$Ho & R.D. Macfarlane & Phys. Rev. &\cite{1963Mac02}& FE & Berkeley & USA &1963 \\
$^{152}$Ho & R.D. Macfarlane & Phys. Rev. &\cite{1963Mac02}& FE & Berkeley & USA &1963 \\
$^{153}$Ho & R.D. Macfarlane & Phys. Rev. &\cite{1963Mac02}& FE & Berkeley & USA &1963 \\
$^{154}$Ho & P. Lagarde & J. Phys. (Paris) &\cite{1966Lag01}& LP & Orsay & France &1966 \\
$^{155}$Ho & A.V. Kalyamin & Sov. At. Energy &\cite{1959Kal01}& SP & Dubna & Russia &1959 \\
$^{156}$Ho & J.W. Mihelich & Phys. Rev. &\cite{1957Mih01}& LP & Oak Ridge & USA &1957 \\
$^{157}$Ho & Z.T. Zhelev & Sov. J. Nucl. Phys. &\cite{1966Zhe01}& SP & Dubna & Russia &1966 \\
$^{158}$Ho & I.S. Dneprovskii & Sov. At. Energy &\cite{1961Dne01}& SP & Dubna & Russia &1961 \\
$^{159}$Ho & K.S. Toth & J. Inorg. Nucl. Chem. &\cite{1958Tot01}& LP & Berkeley & USA &1958 \\
$^{160}$Ho & G. Wilkinson & Phys. Rev. &\cite{1950Wil03}& LP & Berkeley & USA &1950 \\
$^{161}$Ho & T.H. Handley & Phys. Rev. &\cite{1954Han02}& LP & Oak Ridge & USA &1954 \\
$^{162}$Ho & J.W. Mihelich & Phys. Rev. &\cite{1957Mih01}& LP & Oak Ridge & USA &1957 \\
$^{163}$Ho & C.L. Hammer & Phys. Rev. &\cite{1957Ham01}& PN & Ames & USA &1957 \\
$^{164}$Ho & M.L. Pool & Phys. Rev. &\cite{1938Poo02}& LP & Michigan & USA &1938 \\
$^{165}$Ho & F.W. Aston & Nature &\cite{1934Ast04}& MS & Cambridge & UK &1934 \\
$^{166}$Ho & G. Hevesy & Mat.-fys. Medd. &\cite{1936Hev01}& NC & Copenhagen & Denmark &1936 \\
$^{167}$Ho & T.H. Handley & Phys. Rev. &\cite{1955Han01}& LP & Oak Ridge & USA &1955 \\
$^{168}$Ho & R.G. Wille & Phys. Rev. &\cite{1960Wil02}& LP & Arkansas & USA &1960 \\
$^{169}$Ho & K. Miyano & Nucl. Phys. &\cite{1963Miy01}& PN & Tokyo & Japan &1963 \\
$^{170}$Ho & R.G. Wille & Phys. Rev. &\cite{1960Wil02}& LP & Arkansas & USA &1960 \\
$^{171}$Ho & R.M. Chasteler & Z. Phys. &\cite{1989Cha01}& DI & Berkeley & USA &1989 \\
$^{172}$Ho & K. Becker & Nucl. Phys. A &\cite{1991Bec01}& DI & Darmstadt & Germany &1991 \\
 & & & & & & & \\
 & & & & & & & \\
$^{144}$Er & M. Karny & Phys. Rev. Lett. &\cite{2003Kar02}& FE & Oak Ridge & USA &2003 \\
$^{145}$Er & K.S. Vierinen & Phys. Rev. C &\cite{1989Vie01}& FE & Berkeley & USA &1989 \\
$^{146}$Er & K.S. Toth & Phys. Rev. C &\cite{1993Tot01}& FE & Berkeley & USA &1993 \\
$^{147}$Er & G. de Angelis & Z. Phys. A &\cite{1992deA01}& FE & Legnaro & Italy &1992 \\
$^{148}$Er & E. Nolte & Z. Phys. A &\cite{1982Nol02}& FE & Munich & Germany &1982 \\
$^{149}$Er & K.S. Toth & Phys. Rev. C &\cite{1984Tot01}& FE & Berkeley & USA &1984 \\
$^{150}$Er & E. Nolte & Z. Phys. A &\cite{1982Nol02}& FE & Munich & Germany &1982 \\
$^{151}$Er & K.S. Toth & Phys. Rev. C &\cite{1970Tot01}& LP & Oak Ridge & USA &1970 \\
$^{152}$Er & R.D. Macfarlane & Phys. Rev. &\cite{1963Mac01}& FE & Berkeley & USA &1963 \\
$^{153}$Er & R.D. Macfarlane & Phys. Rev. &\cite{1963Mac01}& FE & Berkeley & USA &1963 \\
$^{154}$Er & R.D. Macfarlane & Phys. Rev. &\cite{1963Mac01}& FE & Berkeley & USA &1963 \\
$^{155}$Er & K.S. Toth & Phys. Rev. &\cite{1969Tot01}& LP & Oak Ridge & USA &1969 \\
$^{156}$Er & D. Ward & Phys. Rev. Lett. &\cite{1967War02}& FE & Berkeley & USA &1967 \\
$^{157}$Er & Z.T. Zhelev & Sov. J. Nucl. Phys. &\cite{1966Zhe01}& SP & Dubna & Russia &1966 \\
$^{158}$Er & I.S. Dneprovskii & Sov. At. Energy &\cite{1961Dne01}& SP & Dubna & Russia &1961 \\
$^{159}$Er & A.A. Abdurazakov & Sov. Phys. JETP &\cite{1962Abd01}& SP & Dubna & Russia &1962 \\
$^{160}$Er & M.C. Michel & Phys. Rev. &\cite{1954Mic01}& LP & Berkeley & USA &1954 \\
$^{161}$Er & T.H. Handley & Phys. Rev. &\cite{1954Han02}& LP & Oak Ridge & USA &1954 \\
$^{162}$Er & A.J. Dempster & Phys. Rev. &\cite{1938Dem01}& MS & Chicago & USA &1938 \\
$^{163}$Er & T.H. Handley & Phys. Rev. &\cite{1953Han01}& LP & Oak Ridge & USA &1953 \\
$^{164}$Er & A.J. Dempster & Phys. Rev. &\cite{1938Dem01}& MS & Chicago & USA &1938 \\
$^{165}$Er & F.D.S. Butement & Proc. Phys. Soc. A &\cite{1950But02}& LP & Harwell & UK &1950 \\
$^{166}$Er & F.W. Aston & Nature &\cite{1934Ast04}& MS & Cambridge & UK &1934 \\
$^{167}$Er & F.W. Aston & Nature &\cite{1934Ast04}& MS & Cambridge & UK &1934 \\
$^{168}$Er & F.W. Aston & Nature &\cite{1934Ast04}& MS & Cambridge & UK &1934 \\
$^{169}$Er & A. Bisi & Nuovo Cimento &\cite{1956Bis01}& NC & Harwell & UK &1956 \\
$^{170}$Er & F.W. Aston & Nature &\cite{1934Ast04}& MS & Cambridge & UK &1934 \\
$^{171}$Er & M.L. Pool & Phys. Rev. &\cite{1938Poo02}& LP & Michigan & USA &1938 \\
$^{172}$Er & D.R. Nethaway & Phys. Rev. &\cite{1956Net01}& NC & Berkeley & USA &1956 \\
$^{173}$Er & V. Pursiheimo & Z. Phys. &\cite{1972Pur01}& LP & Helsinki & Finland &1972 \\

$^{174}$Er & R.M. Chasteler & Z. Phys. &\cite{1989Cha01}& DI & Berkeley & USA &1989 \\
$^{175}$Er & X. Zhang & Z. Phys. &\cite{1996Zha01}& LP & Lanzhou & China &1996 \\
 & & & & & & & \\
 & & & & & & & \\
$^{145}$Tm& J.C. Batchelder & Phys. Rev. C &\cite{1998Bat01}& FE & Oak Ridge & USA &1998 \\
$^{146}$Tm & K. Livingston & Phys. Lett. B &\cite{1993Liv01}& FE & Darmstadt & Germany &1993 \\
$^{147}$Tm & O. Klepper & Z. Phys. A &\cite{1982Kle01}& FE & Darmstadt & Germany &1982 \\
$^{148}$Tm & E. Nolte & Z. Phys. A &\cite{1982Nol02}& FE & Darmstadt & Germany &1982 \\
$^{149}$Tm & K.S. Toth & Phys. Rev. C &\cite{1987Tot02}& FE & Berkeley & USA &1987 \\
$^{150}$Tm & E. Nolte & Z. Phys. A &\cite{1982Nol02}& FE & Darmstadt & Germany &1982 \\
$^{151}$Tm & H. Helppi & Phys. Lett. B &\cite{1982Hel01}& FE & Argonne & USA &1982 \\
$^{152}$Tm & C.F. Liang & Z. Phys. A &\cite{1980Lia01}& SP & Orsay & France &1980 \\
$^{153}$Tm & R.D. Macfarlane & Phys. Rev. &\cite{1964Mac01}& FE & Berkeley & USA &1964 \\
$^{154}$Tm & R.D. Macfarlane & Phys. Rev. &\cite{1964Mac01}& FE & Berkeley & USA &1964 \\
$^{155}$Tm & K.S. Toth & Phys. Rev. C &\cite{1971Tot01}& FE & Oak Ridge & USA &1971 \\
$^{156}$Tm & K.S. Toth & Phys. Rev. C &\cite{1971Tot01}& FE & Oak Ridge & USA &1971 \\
$^{157}$Tm & J.C. Putaux& Nucl. Instrum. Meth. &\cite{1974Put01}& LP & Orsay & France &1974 \\
$^{158}$Tm & F.W.N. de Boer& Radiochim. Acta &\cite{1970deB01}& LP & Amsterdam & Netherlands &1970 \\
$^{159}$Tm & C. Ekstrom & Nucl. Phys. A &\cite{1971Eks01}& LP & Uppsala & Sweden &1971 \\
$^{160}$Tm & F.W.N. de Boer& Radiochim. Acta &\cite{1970deB01}& LP & Amsterdam & Netherlands &1970 \\
$^{161}$Tm & B. Harmatz & Phys. Rev. &\cite{1959Har01}& LP & Oak Ridge & USA &1959 \\
$^{162}$Tm & A. Abdumalikov & Phys. Lett. &\cite{1963Abd01}& SP & Dubna & Russia &1963 \\
$^{163}$Tm & B. Harmatz & Phys. Rev. &\cite{1959Har01}& LP & Oak Ridge & USA &1959 \\
$^{164}$Tm & A. Abdurazakov & Nucl. Phys. &\cite{1960Abd01}& SP & Dubna & Russia &1960 \\
$^{165}$Tm & T.H. Handley & Phys. Rev. &\cite{1953Han01}& LP & Oak Ridge & USA &1953 \\
$^{166}$Tm & G. Wilkinson & Phys. Rev. &\cite{1948Wil02}& LP & Berkeley & USA &1948 \\
$^{167}$Tm & G. Wilkinson & Phys. Rev. &\cite{1948Wil02}& LP & Berkeley & USA &1948 \\
$^{168}$Tm & G. Wilkinson & Phys. Rev. &\cite{1949Wil03}& LP & Berkeley & USA &1949 \\
$^{169}$Tm & F.W. Aston & Nature &\cite{1934Ast04}& MS & Cambridge & UK &1934 \\
$^{170}$Tm & E. Neuninger & Wiener Akad. Anzeiger &\cite{1936Neu01}& NC & Wien & Austria &1936 \\
$^{171}$Tm & S. DeBenedetti & Phys. Rev. & \cite{1948DeB01} & NC & Oak Ridge & USA &1948 \\
$^{172}$Tm & D.R. Nethaway & Phys. Rev. &\cite{1956Net01}& NC & Berkeley & USA &1956 \\
$^{173}$Tm & T. Kuroyanagi & J. Phys. Soc. Japan &\cite{1961Kur01}& PN & Tohoku & Japan &1961 \\
$^{174}$Tm & R.G. Wille & Phys. Rev. &\cite{1960Wil02}& LP & Arkansas & USA &1960 \\
$^{175}$Tm & T. Kuroyanagi & J. Phys. Soc. Japan &\cite{1961Kur01}& PN & Tohoku & Japan &1961 \\
$^{176}$Tm & K. Takahashi & J. Phys. Soc. Japan &\cite{1961Tak01}& LP & Tokyo & Japan &1961 \\
$^{177}$Tm & K. Rykaczewski & Nucl. Phys. A &\cite{1989Ryk01}& DI & Darmstadt & Germany &1989 \\
 & & & & & & & \\
 & & & & & & & \\
$^{149}$Yb& S.-W. Xu & Eur. Phys. J. A &\cite{2001Xu01}& FE & Lanzhou & China &2001 \\
$^{150}$Yb & & & & & & & \\
$^{151}$Yb & P. Kleinheinz & Z. Phys. A &\cite{1985Kle01}& FE & Darmstadt & Germany &1985 \\
$^{152}$Yb & E. Nolte & Z. Phys. A &\cite{1982Nol01}& FE & Munich & Germany &1982 \\
$^{153}$Yb & E. Hagberg & Nucl. Phys. A &\cite{1977Hag02}& SP & CERN & Switzerland &1977 \\
$^{154}$Yb & R.D. Macfarlane & Phys. Rev. &\cite{1964Mac01}& FE & Berkeley & USA &1964 \\
$^{155}$Yb & R.D. Macfarlane & Phys. Rev. &\cite{1964Mac01}& FE & Berkeley & USA &1964 \\
$^{156}$Yb & K.S. Toth & Phys. Rev. C &\cite{1970Tot01}& LP & Oak Ridge & USA &1970 \\
$^{157}$Yb & K.S. Toth & Phys. Rev. C &\cite{1970Tot01}& LP & Oak Ridge & USA &1970 \\
$^{158}$Yb & D. Ward & Phys. Rev. Lett. &\cite{1967War02}& FE & Berkeley & USA &1967 \\
$^{159}$Yb & W. Trautmann & Phys. Rev. Lett. &\cite{1975Tra01}& FE & Munich & Germany &1975 \\
$^{160}$Yb & D. Ward & Phys. Rev. Lett. &\cite{1967War02}& FE & Berkeley & USA &1967 \\
$^{161}$Yb & A. Latuszynski & Nukleonika &\cite{1974Lat02}& SP & Dubna & Russia &1974 \\
$^{162}$Yb & A. Abdumalikov & Phys. Lett. &\cite{1963Abd01}& SP & Dubna & Russia &1963 \\
$^{163}$Yb & P. Paris & Compt. Rend. Acad. Sci. &\cite{1967Par01}& LP & Orsay & France &1967 \\
$^{164}$Yb & F.D.S. Butement & J. Inorg. Nucl. Chem. &\cite{1960But02}& SP & Liverpool & UK &1960 \\
$^{165}$Yb & P. Paris & Compt. Rend. Acad. Sci. &\cite{1964Par01}& LP & Orsay & France &1964 \\
$^{166}$Yb & M.C. Michel & Phys. Rev. &\cite{1954Mic01}& LP & Berkeley & USA &1954 \\
$^{167}$Yb & T.H. Handley & Phys. Rev. &\cite{1954Han01}& LP & Oak Ridge & USA &1954 \\
$^{168}$Yb & A.J. Dempster & Phys. Rev. &\cite{1938Dem01}& MS & Chicago & USA &1938 \\
$^{169}$Yb & W. Bothe & Z. Naturforsch. &\cite{1946Bot01}& NC & Berlin & Germany &1946 \\
$^{170}$Yb & A.J. Dempster & Phys. Rev. &\cite{1938Dem01}& MS & Chicago & USA &1938 \\
$^{171}$Yb & F.W. Aston & Nature &\cite{1934Ast04}& MS & Cambridge & UK &1934 \\
$^{172}$Yb & F.W. Aston & Nature &\cite{1934Ast04}& MS & Cambridge & UK &1934 \\
$^{173}$Yb & F.W. Aston & Nature &\cite{1934Ast04}& MS & Cambridge & UK &1934 \\
$^{174}$Yb & F.W. Aston & Nature &\cite{1934Ast04}& MS & Cambridge & UK &1934 \\
$^{175}$Yb & H. Atterling & Arkiv Mat. Astron. Fysik&\cite{1945Att01}& NC & Stockholm & Sweden &1945 \\
$^{176}$Yb & F.W. Aston & Nature &\cite{1934Ast04}& MS & Cambridge & UK &1934 \\
$^{177}$Yb & H. Atterling & Arkiv Mat. Astron. Fysik&\cite{1945Att01}& NC & Stockholm & Sweden &1945 \\
$^{178}$Yb & C.J. Orth & Phys. Rev. C &\cite{1973Ort01}& LP & Los Alamos & USA &1973 \\
$^{179}$Yb & R. Kirchner & Nucl. Phys. A &\cite{1982Kir01}& DI & Darmstadt & Germany &1982 \\
$^{180}$Yb & E. Runte & Z. Phys. A &\cite{1987Run01}& DI & Darmstadt & Germany &1987 \\
 \\
\end{longtable}

\end{document}